\documentclass[12pt]{iopart}

\usepackage{mathrsfs}         
\usepackage{epsfig}
\usepackage{graphicx}
\usepackage{color}
\newcommand{\rev}[1]{{#1}}
\newcommand{\rv}[1]{{#1}}

\newcommand{\revv}[1]{{#1}}

\newcommand{\cor}[1]{{#1}}

\begin{document}

\title[Influence of interactions on the anomalous quantum Hall effect]{Influence of interactions on the anomalous quantum Hall effect}

\author{C.X. Zhang}

\address{Physics Department, Ariel University, Ariel 40700, Israel}

\author{M.A. Zubkov} 

\address{Physics Department, Ariel University, Ariel 40700, Israel}

\vspace{10pt}

\begin{abstract}
The anomalous quantum Hall conductivity in the 2+1D topological insulators in the absence of interactions may be expressed as the topological invariant composed of the two - point Green function. For the noninteracting system this expression is the alternative way to represent the TKNN invariant. It is widely believed that in the presence of interactions the Hall conductivity is given by the same expression, in which the noninteracting two - point Green function is substituted by the complete two - point Green function with the interactions taken into account. However, the proof of this statement has not been given so far. In the present paper we give such a proof in the framework of the particular  tight - binding models of the $2+1$ D topological insulator. Besides, we extend our consideration to the $3+1$ D Weyl semimetals. It was known previously that with the interactions neglected the Hall conductivity in those systems is expressed through the two - point Green function in the way similar to that of the $2+1$ D topological insulators. Again, the influence of interactions on this expression has not been investigated previously. We consider this problem within the framework of the particular $3+1$D model of Weyl semimetal in the presence of the contact four - fermion interactions and Coulomb interactions. We prove (up to the one - loop approximation), that the Hall conductivity is given by the same expression as in the noninteracting case, in which the noninteracting Green function is substituted by the complete two - point Green function with the interactions included. Basing on the obtained expressions we discuss the topological phase transitions accompanied by the change of Hall conductivity.
\end{abstract}

%
%
%
%
%



\section{Introduction}
\label{SectIntro}

The Anomalous Quantum Hall effect (AQHE) is related to the Hall current
which appears in the absence of external magnetic field.
The understanding of AQHE involves the concepts of topology and geometry \cite{nagaosa}. 
The topological invariant responsible for the AQHE
in the ideal two space-dimensional (2d) non-interacting condensed matter systems has been proposed in \cite{TTKN}.
It is called now the TKNN invariant and is given by the integral of Berry curvature
over the occupied electronic states \cite{Fradkin,Tong:2016kpv,Hatsugai}.
An extension of this approach to the three space-dimensional (3d) topological insulators
was considered, in particular, in \cite{Hall3DTI}. It is widely believed, that the introduction
of weak interactions does not affect Hall conductivity. Therefore, it is important to express it
through the quantities defined within the interacting theory. The two - point Green function is such a quantity.
The expression of the AQHE conductivity for the $2+1$ D (2d for short) systems
through the topological invariant composed of the Green functions has been proposed
in \cite{Matsuyama:1986us,Imai:1990zz,Volovik0} (see also Chapter 21.2.1 in \cite{Volovik2003}).
The extension of this construction to various 3d systems has been given in \cite{Z2016_1},
where, in particular, the description of the AQHE in topological Weyl semimetals \cite{semimetal_effects10,semimetal_effects11,semimetal_effects12,semimetal_effects13,Zyuzin:2012tv,tewary}) was given.
We notice here also the discussion of the similar topological invariants
in \cite{Gurarie2011,EssinGurarie2011}. It is widely believed that
the AQHE conductivity is given by the expressions of  \cite{Matsuyama:1986us,Volovik0,Volovik2003,Z2016_1}
expressed through the two - point Green functions of interacting systems.
However, there is still no direct proof that the other interaction contributions to the AQHE are absent.

In a recent paper \cite{ZW2019} written with the participation of one of the present authors,
the construction of \cite{Z2016_1} was further extended to the non - homogeneous systems.
This paper \cite{ZW2019} indicated that the Hall conductivity is still given by
the expression through the two - point Green functions presented in \cite{Z2016_1},
in the presence of varying magnetic field.
Its construction allows to give a relatively simple alternative proof
that weak disorder does not affect the AQHE conductivity.
This work complements the previous consideration of the role of disorder
in the quantum Hall effect (QHE) \cite{Fradkin,TKNN2,QHEr,Tong:2016kpv,Zheng+Ando2002}.
The absence of corrections (due to weak Coulomb interactions) to the QHE
in the disordered ferromagnetic metal was shown in \cite{nocorrectionstoQHE}.
The effects of interactions between the electrons to the QHE in
2DEG has been discussed long time ago (see, for example,
\cite{Altshuler0,Altshuler} and references therein), and the exactness of
Hall plateaus were discussed.
In \cite{Ishikawa:2003tu,Imai:1990zz} (see also references therein) the Hall conductivity is studied on the basis of the Ward-Takahashi identity with the electron-electron interactions taken into account. In principle, the results of these papers may be extended to the intrinsic Anomalous Quantum Hall effect. However, in our present paper we consider this problem using a different technique.
It is also worth mentioning, that the strong interactions are able to lead the fermionic system in the presence of external magnetic field to the fractional QHE phases \cite{Tong:2016kpv}.

The similar phenomena for the AQHE existing without magnetic field has been discussed as well (see \cite{FAQHE} and references therein).
Interaction effects in the 2d topological insulators were discussed, for example,
in \cite{2DTI_corr} (see also references therein).
Recently the effects of electron-electron interactions were investigated in graphene-like systems,
and the renormalization of Fermi velocity was studied, taking the Coulomb interactions into account \cite{Tang2018_Science}.
The corrections due to the simple on - site interactions to the AQHE
in the 2D insulators were discussed, for example, in \cite{corr_2d,corr_2d_2}.
In \cite{AQHE_no_corr} the corrections due to interactions to the AQHE
in the Weyl semimetals were considered. The authors of \cite{AQHE_no_corr}
restricted themselves to the simple Hubbard interactions on the lattice model. The interaction corrections in Weyl seimimetals
were also discussed in \cite{corr_WSM1,corr_WSM2} without any relation to the AQHE.
To the best of our knowledge, the corrections to Hall conductivity
due to the most relevant Coulomb interactions were not discussed both for
the 2d topological insulators (where the AQHE exists due to the nontrivial band topology)
and for 3d Weyl semimetals
\footnote{It is worth mentioning that in the framework of the relativistic quantum field theory the related question has been considered. Namely, in the $2+1$ D QED the integration over fermions leads to the appearance of the term in effective action proportional to the Chern - Simons term. This phenomenon known also as parity anomaly. The coefficient standing in front of the Chern - Simons terms is proportional to the Hall conductivity in the given system. It has been proved in \cite{parity_anomaly} that this coefficient has contributions only in the one - loop approximation while the contributions of the higher orders of perturbation theory vanish. The one loop expression is proportional to the sign of the fermion mass summed up over the existing $2+1$ D relativistic fermions. The obtained result is known as the non - renormalization of parity anomaly. However, this non - renormalization in the framework of relativistic $2+1$ D QED does not prove the conjecture considered in the present paper. First of all, the systems considered in \cite{parity_anomaly} are much more simple than those that are discussed in our paper. (Therefore, in the simple $2+1$ D QED the complicated pattern of topological transition discussed in the present paper is not observed.) Second, in \cite{parity_anomaly} it was not proved that the Hall conductivity is expressed through the interacting Green functions in the same way as in the absence of interactions.}.

In the present paper we report the results of the investigation that partially fills this gap.
Namely, we discuss here the tight - binding models in 2d and 3d
that correspond to the 2d topological insulators and 3d Weyl semimetals.
The tight - binding models of Weyl semimetals considered here
are actually those of \cite{tb1,tb2,tb3,tb4,tb5}
while the models of the 2d topological insulators are those considered previously,
for example, in \cite{Z2016_1,tb2d,tb2d2,tb2d3,corr_2d}.
Those models give qualitative description of the really existing systems.
We will show here, that both the four - fermion interactions and the interactions
due to exchange by scalar bosons (including the Coulomb interactions)
do not give any corrections to the AQHE conductivities of \cite{Z2016_1}
expressed through the (interacting) Green functions,
at least in the one - loop approximation. We also have found,
that the sufficiently strong interactions may lead to the topological phase transitions.
For various other properties of Weyl semimetals we refer to
\cite{V1,V2,V3,V4,V5,V6,V7,S1,S2,S3,S4,S5,S6,S7,B1,B2,B3,B4,B5,B6,G1,G2,G3}.

On the technical side we use the version of Wigner - Weyl technique \cite{1,2,berezin,6} adapted in \cite{Z2016_1} to the lattice models of solid state physics combined with the ordinary perturbation theory (see also \cite{SZ2018}). We expect that the  obtained result may be extended further to the consideration of the other non - dissipative transport phenomena (for the review of the latter see \cite{Landsteiner:2012kd} and references therein). \revv{The main advantage of the Wigner - Weyl formalism is that it allows to express the Hall conductivity directly through the interacting two - point Green function. This expression is the topological invariant for the topological insulators and as such it is robust to the weak modification of interactions. For the Weyl semimetals, the expression for the Hall conductivity obtained in this way inherits the algebraic structure of the expression for the topological insulators, is expressed through the two - point interacting Green function, and does not contain the higher order ones.}

 The paper is organized as follows.
 In Sect. II we consider the tight - binding model of the 2d topological insulator
 with the four - fermion interaction,
Yukawa or Coulomb interactions among the electrons.
In Sect. III, we study the 3d Weyl semimetals in the presence of these interactions.
 In Sect. IV, we end with the conclusions.

\section{AQHE in the $2+1$ D tight - binding model}
\label{SectHall}

\subsection{Neglecting interactions}
Let us start from the 2+1 D lattice model of the non-interacting fermions
\begin{eqnarray}\label{lagrangian}
S_{0} = \int d\tau \sum_{{\bf x,x'}}\bar{\psi}_{\bf x'}\Big(i(i \partial_{\tau} - A_3(-i\tau,{\bf x}))\delta_{\bf x,x'}
- i{\cal D}_{\bf x,x'}\Big)\psi_{\bf x}
\end{eqnarray}
where $\bar{\psi}$ is the Hermitian conjugation of $\psi$, i.e. $\bar{\psi}=\psi^{\dagger}$, $\tau$ is imaginary time ($t=-i\tau$), and
\begin{eqnarray}\label{lattice_difference}
{\cal D}_{\bf x,x'}= &-&\frac{i}{2} \sum_{i=1,2} [(1+\sigma^i)\delta_{x+e_i,x'}e^{iA_{x+e_i,x}}
                           + (1-\sigma^i)\delta_{x-e_i,x'}e^{iA_{x-e_i,x}}] \sigma_3     \nonumber\\
                   &+& i(m+2)\delta_{\bf x,x'}\sigma_3
\end{eqnarray}
where $A_{u,v}= \int^{u}_{v} A \cdot ds$.
$A_3$ is expressed through the external electric potential $\phi(t,{\bf x})$ as $A_3  =  -i \phi(-i \tau, {\bf x} )$. Space coordinates $\bf x$ are discrete while the values of $\tau$ are continuous. We denote the Euclidean three - momentum by $p = (\omega, {\bf p})$. In 3D Euclidean coordinate space, a point is denoted by $x = (\tau, {\bf x})$, and the Euclidean 3 - potential is $A = (-i \phi, {\bf A})$.

{Fermionic field in momentum space can be defined through
the Fourier transform
\begin{eqnarray}\label{Fourier}
\psi(p)=\sum_{\bf x} e^{-ipx}\int \psi_{\bf x,\tau} e^{-i\omega \tau} d\tau,
\end{eqnarray}
which is periodic with the perion $2\pi$ as a function of $\bf p$.}
Let us define
$Q(p,x)=Q(\omega,{\bf p},\tau,{\bf x})=i(\omega - \phi(i\tau,{\bf x})) - H({\bf p} - {\bf A}(i\tau,{\bf x}))$.
Then the free Green function in momentum space  $\tilde{G}_0 (p_1,p_2)$
is defined as
\begin{eqnarray}\label{Green_p}
\tilde{G}_{0}(p_2,p_1) &=& \frac{1}{Z_0}
\int \frac{D\bar{\psi}D{\psi}}{(2\pi)^3}\,
       \bar{\psi}(p_2)\psi(p_1)
      e^{\int \frac{d^3 p}{(2\pi)^3} \bar{\psi}(p) Q(p,i\partial_p)\psi(p)},
\end{eqnarray}
{ where $\int d^3 p =\int^{\infty}_{-\infty} d\omega \int^{\pi}_{-\pi} d^2\bf{p}$.
It satisfies  equation}
\begin {eqnarray}\label{Green_k_equ}
Q(p_1,i\partial_{p_1})\tilde{G}_0(p_1,p_2)=\delta^3 (p_1-p_2).
\end{eqnarray}
The free Green function in coordinate space $G_0(x_1,x_2)$
is related to $\tilde{G}_0$ by the Fourier transformation
\begin {eqnarray}\label{Green_pp2xx}
G_0(x_1,x_2)=\int \frac{d^3 p_1}{(2\pi)^{3/2}} \int \frac{d^3 p_2}{(2\pi)^{3/2}}
 e^{ip_1 x_1} \tilde{G}_0(p_1,p_2) e^{-ip_2 x_2}
\end{eqnarray}
All components of variable $x_i$ in the above equation may take continuous values.
Similar to Eq.(\ref{Green_k_equ}),
{ if $\bf x'_1$ and $\bf x'_2$ take discrete (integer) values,
then $G_0(x_1,x_2)$ satisfies
\begin {eqnarray}\label{Green_x_equ}
Q(-i\partial_{x_1},x_1) G_0(x_1,x'_2) \Big|_{x_1=(\tau_1,\bf x'_1)}=\delta (\tau_1-\tau_2)\delta_{\bf x'_1,\bf x'_2}.
\end{eqnarray}   } 
This may be proved directly, as follows. Let us represent function $Q(-i\partial_{x_1},x_1)$ as a series in powers of $-i\partial_{x_1}$ and $x_1$. We order those operators in such a way that the powers of $-i\partial_{x_1}$ are right to the powers of $x_1$. Then we represent
\begin {eqnarray}\label{Green_x_equ_proofA}
& & Q(-i\partial_{x_1},x_1) G_0(x_1,x'_2)  \nonumber \\
&=& \int \frac{d^3 p_1 d^3 p_2}{(2\pi)^{3}}
 [Q(-i\partial_{x_1},x_1) e^{ip_1 x_1}] \tilde{G}_0(p_1,p_2) e^{-ip_2 x'_2}      \nonumber\\
&=& \int \frac{d^3 p_1 d^3 p_2}{(2\pi)^{3}}
[e^{ip_1 x_1} Q(p_1,x_1) ] \tilde{G}_0(p_1,p_2) e^{-ip_2 x'_2}\nonumber\\
&=& \int \frac{d^3 p_1 d^3 p_2}{(2\pi)^{3}}
[e^{ip_1 x_1} Q(p_1,-i\overleftarrow{\partial}_{p_1}) ] \tilde{G}_0(p_1,p_2) e^{-ip_2 x'_2}.
\end{eqnarray}
{Integration by parts
will be applied to the last line of Eq.(\ref{Green_x_equ_proofA}),
and the additional boundary term can be omitted,
after taking the limit ${\bf x_1} \rightarrow {\bf x'_1}$,
where ${\bf x'_1}$ take integer values.
Then taking into account Eq.(\ref{Green_k_equ}),
one finds that  }
\begin {eqnarray}\label{Green_x_equ_proofB}
&& Q(-i\partial_{x_1},x_1) G_0(x_1,x'_2) \Big|_{x_1=(\tau_1,\bf x'_1)}   \nonumber \\
&=& \int \frac{d^3 p_1 d^3 p_2}{(2\pi)^{3}}
e^{ip_1 x'_1} [Q(p_1,i\partial_{p_1}) \tilde{G}_0(p_1,p_2) ]  e^{-ip_2 x'_2}    \nonumber  \\
&=& \int \frac{d^3 p_1 d^3 p_2}{(2\pi)^{3}}
e^{ip_1 x'_1}   \delta(p_1-p_2) e^{-ip_2 x'_2}    \nonumber  \\
&=& \delta (\tau_1-\tau_2)\delta_{\bf{x'_1},\bf{x'_2}}.
\end{eqnarray}

Applying Wigner transformation, one obtains (assuming that the field $A$ is slowly varying,
i.e. when its variations on the distances of the order of the lattice spacing may be neglected)
the Groenewold equation
\begin {eqnarray}\label{Green0_w_equ}
Q_W(x,p)\star G_{0,W}(x,p) =1
\end{eqnarray}
where $Q_W(x,p)$ and $G_{0,W}$ are the Wigner transformations of $Q$ and $G_0$, respectively:
\begin{eqnarray}
{G}_{0,W}(x,p)& = & \int d^3q e^{ix q} \tilde{G}_0({p+q/2}, {p-q/2})\label{GW_def}\\
{Q}_{W}(x,p)& = & \int d^3q e^{ix q} \tilde{Q}({p+q/2}, {p-q/2}), \nonumber
\end{eqnarray}
where
\begin{equation}
\tilde{Q}(p_1,p_2) \equiv  \int d^3k \delta^{(3)}(p_1-k) {Q}(k,i\partial_{k}) \delta^{(3)}(p_2 - k)\nonumber
\end{equation}
represents the matrix elements of operator $\hat{Q}$. Star product $\star$ is the operation
$\star=e^{i\tilde{\Delta}/2}$, with
 $\tilde{\Delta} =\overleftarrow{\partial}_x\overrightarrow{\partial}_p-\overleftarrow{\partial}_p\overrightarrow{\partial}_x$.
The gradient expansion further gives $G_{0,W}=G^{(0)}_{0,W}+G^{(1)}_{0,W}+...$
with $G^{(n)}_{0,W} \sim O(\partial^n_x)$.
$G^{(0)}_{0,W}$ is given by $G^{(0)}_{0,W}(x,p)=g(p-{\cal A}(x))$ \cite{SZ2018,FZ2019}, in which
$
g(p) =[i\omega  - H({\bf p} )]^{-1},
$
and ($\mu=1,2$)
\begin{equation}\label{calAs_a}
{\cal A}_\mu({\bf x}) =  \int \big[\frac{\sin(k_{\mu}/2)}{k_{\mu}/2}\tilde A_\mu({\bf k})e^{i{\bf kx}}+c.c. \big]dk
\end{equation}
that is
\begin{eqnarray} \label{calAs_b}
{\cal A}_1({\bf x}) & = & \int_{{\bf x} - {\bf e}_1/2}^{{\bf x} + {\bf e}_1/2} A_1(y_1,x_2)dy_1 \nonumber\\
{\cal A}_2({\bf x}) & = & \int_{{\bf x} - {\bf e}_2/2}^{{\bf x} + {\bf e}_2/2} A_2(x_1,y_2)dy_2
\end{eqnarray}
where ${\bf e}_{\mu}$ is the unit lattice vector directed along the $\mu$ - th axis. (The original electromagnetic field itself may be represented in the form:
${A}_\mu({\bf x}) =  \int \big[\tilde A_\mu({\bf k})e^{i{\bf kx}}+c.c. \big]dk $.) For the slowly varying electromagnetic fields we may substitute $\cal A$ by $A$, which will be done further.
$G^{(1)}_{0,W}$ is given by \cite{SZ2018,Zu1,Zu2,Zu3,Zu4,Zu5,ter}
\begin{equation}\label{G0(1)}
G_{0,W}^{(1)}(x,p) = \cor{+}\frac{i}{2}  G_{0,W}^{(0)}  \frac{\partial [G_{0,W}^{(0)}]^{-1}}{\partial p_i}
                               G_{0,W}^{(0)}  \frac{\partial [G_{0,W}^{(0)}]^{-1}}{\partial p_j}
                               G_{0,W}^{(0)}  F_{ij}
\end{equation}

Electric current can be considered as the linear response to the external field,
i.e. $\delta {\rm log}\,Z=J^k(x)\,\delta A_k(x)$, with $Z$ the partition function.
The electric current (along the $x_k$-axis) is given by
\begin{eqnarray}\label{current_2D}
J^k(x)& = & - \int \frac{d^3 p}{(2\pi)^3}  Tr G_{0,W}(x,p) \frac{\partial}{\partial p_k}[G^{(0)}_{0,W}(x,p)]^{-1}.
\end{eqnarray}
Corresponding to $G=G^{(0)}+G^{(1)}+...$,
$J_k$ is expanded into $J_k=J_k^{(0)}+J_k^{(1)}+...$.
The term linear in the field strength is written as
\begin{eqnarray}
J_k^{(1)}(x)  &= &\cor{-} \frac{1}{4\pi}\epsilon_{ijk} {\cal M}_{} F_{ij} (x), \label{j2d}\\
 {\cal M} &=& \rv{ \frac{i}{3!\,4\pi^2}\,\epsilon_{ijk} \int_{} \,{\rm Tr}\, d^3p  \, \Big[ {\cal G}^{-1} \partial_{p_i} {\cal G} \partial_{p_j} {\cal G}^{-1} \partial_{p_k} {\cal G}\Big]}  \nonumber
\end{eqnarray}
in which the Green function ${\cal G}$ satisfies ${\cal G}^{-1} = i \omega - H({\bf p})$,
with $H$ the one - particle Hamiltonian.
Here $F_{ij}$ is the Euclidean field strength
$F_{ij} = \partial_i A_j  - \partial_j A_i$.  In the present paper we define the components  $A_k = A^k$ for $k = 1,2$  as equal to the space components of real external electromagnetic potential $\bf A$ in Minkowski space - time. Correspondingly, $A_3 = - A^3 = -i A^0$, where $A^0$ is the external electric potential. In order to make the present paper self - contained we give the derivation of Eq. (\ref{j2d}) in Appendix A.
The generalization to the case of the $3+1$ D models is straightforward. It is worth mentioning, that the derivation of Eq. (\ref{j2d}) requires that the field $A$ does not vary fast, i.e. its variation on the distance of the order of lattice spacing may be neglected.

We suppose, that the fermions are gapped and the Green function ${\cal G}({ p})$
depends on the three - vector ${p} = (p_1,p_2,p_3)$ of Euclidean momentum
(the third component of vector corresponds to imaginary time).
In order to obtain expression for the Hall current
let us introduce into  Eq. (\ref{j2d}) the external electric field
${\bf E} = (E_1,E_2)$ as $A_{3k} = -i E_k$ .
This leads to the following expression for the Hall current
\begin{equation}
{j}^k_{Hall} = \cor{-}\frac{1}{2\pi}\,{\cal N}\,\epsilon^{ki}E_i,\label{HALLj}
\end{equation}
where the topological invariant denoted by ${\cal N}$ is to be calculated for the original system with vanishing background gauge field:
\begin{eqnarray}
{\cal N} &=&  \rv{-\frac{1}{24 \pi^2}} {\rm Tr}\, \int_{} {\cal G}^{-1} d {\cal G} \wedge d {\cal G}^{-1} \wedge d {\cal G}\label{N3A}
\end{eqnarray}
Eq. (\ref{N3A}) defines the topological invariant (this is proved, in particular, in Appendix B of  \cite{Z2016_1}). Recall, that for the given lattice model $\cal G$ is the Green function in momentum space, i.e. the Fourier transformation of the two point Green function in coordinate space (it is assumed that the original model without external gauge field is translation invariant).

The value of ${\cal N}$ is computable.
For  example, we consider the Green function of the form ${\cal G}^{-1} = i \omega - H({\bf p})$ with the Hamiltonian
\begin{equation}
H = {\rm sin}\,p_1\, \sigma^2 - {\rm sin}\, p_2 \, \sigma^1 - (m + \sum_{i=1,2}(1-{\rm cos}\,p_i)) \, \sigma^3. \label{Ham0}
\end{equation}
For $m \in (-2,0)$ we have ${\cal N} = 1$,
while  ${\cal N}  = -1 $ for $m\in (-4,-2)$,
and ${\cal N}  = 0 $ for $m\in (-\infty,-4)\cup (0,\infty)$.
The calculation is given in Appendix B.

\subsection{The four - fermion interaction}

We add to the model discussed above the four - fermion interaction term, which gives the Euclidean action
\begin{eqnarray}\label{lagrangian}
S_{\lambda}= S_0
           +\frac{\lambda}{2}  \int d\tau \sum_{{\bf x}}(\bar{\psi}(\tau,{\bf x})\psi(\tau, {\bf x}))^2,
\end{eqnarray}
Therefore, the partition function and the two-point Green function
are given by
\begin {eqnarray}\label{partition}
Z_{\lambda} = \int\rv{ D\bar{\psi}D{\psi} e^{S_{\lambda}}}
\end{eqnarray}
and
\begin {eqnarray}\label{Green}
G=G_{\lambda}(x_1,x_2) =\frac{1}{Z_{\lambda}}
\int D\bar{\psi}D{\psi}\, \bar{\psi}(x_1) {\psi}(x_2) e^{S_{\lambda}},
\end{eqnarray}

\begin{figure}[h]
	\centering  %
	\includegraphics[width=0.5\linewidth]{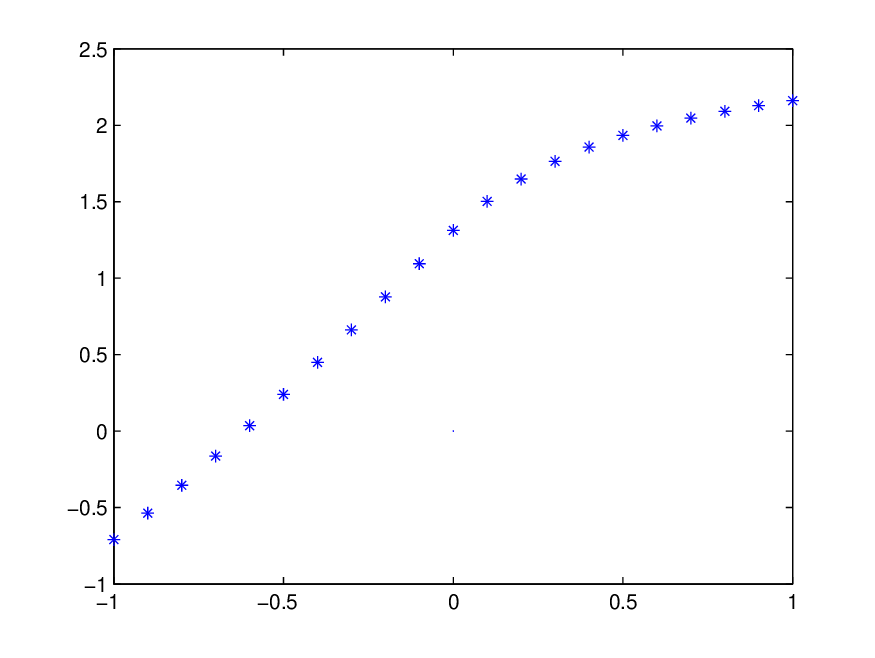}  %
	\caption{The dependence of $\xi$ on mass parameter.
             The horizontal axis represents the dimensionless mass parameter,
             i.e. the physical value of $m$ multiplied by the lattice constant.}  %
	\label{fig.1}   %
\end{figure}

In the presence of the 4 - fermion interaction and the external field $A_{\mu}$, the Green function $G(x_1,x_2)$ can be expressed as
\begin {eqnarray}\label{Green_tower}
G(x_1,x_2) &=& G_{0}(x_1,x_2) \nonumber\\
&& + \lambda \int d\tau_y \sum_{\bf y} G_{0}(x_1,y)H_{0}(y)G_{0}(y,x_2)   \nonumber\\
&& - \lambda \int d\tau_y \sum_{\bf y} ({\rm Tr}\, H_{0}(y)) G_{0}(x_1,y)G_{0}(y,x_2)    \nonumber\\
&& +O(\lambda^2) \nonumber\\
&=& G_{0} +  W*G
\end{eqnarray}
according to the Feynman diagrams, where { $H_{0}(y)= G_{0}(y,y)$   }.
   For an arbitary function $f(x_1,x_2)$, the convolution $W*f$ is defined as
$\lambda\int d\tau_y \sum_{\bf y} G_{0}(x_1,y)\Xi_{0}(y)f(y,x_2)$, with
{ $\Xi_{0}(y)=H_{0}(y)-{\rm Tr}\, H_{0}(y)$}. From Eq.(\ref{Green_tower}), one obtains $(1-W)*G=G_0$,
in which corrections in $\lambda^2$ order and higher have been neglected. Applying $Q$ to both sides, we get
\begin {eqnarray}\label{Green_equ2}
Q(i\partial_x,x)G(x,y)- \lambda \Xi_0(x)G(x,y) =\delta(x_3-y_3)\delta_{\bf xy}
\end{eqnarray}
Wigner transformation gives
\begin {eqnarray}\label{Green_equ3}
(Q_W(x,p)- \lambda \Xi_0(x))\star G_W(x,p) =1
\end{eqnarray}
where $G_W$  is the Wigner transformation of $G$.
In the presence of the 4 - fermion interaction, the gradient expansion of the Green function is similar:
$G_{W}(x,p)=G^{(0)}_{W}+G^{(1)}_{W}+...$,
where $G^{(n)}_{W} \sim O(\partial_x^{n})$.
The leading order term $G^{(0)}_{W}(R,p)$ contains the correction from the 4 - fermion interaction
\begin {eqnarray}\label{Green_W_express}
G^{(0)}_{W}(x,p) =[i(\omega - A_3(x)) - H({\bf p} - {A}(x))-\lambda \Xi^{(0)}_0(x)]^{-1}, \nonumber
\end{eqnarray}
in which
\begin {eqnarray}\label{Xi_00}
\Xi^{(0)}_0 (x) &=& \int \frac{d^3 p}{(2\pi)^3} G^{(0)}_{0,W}(x,p) \\ \nonumber
              &=& \int\frac{d^2 {\bf p}}{(2\pi)^2} \int \frac{d \omega}{2\pi}
              \frac{-H({\bf p}- {A}(x))}{\omega^2  + H({\bf p} - {A}(x) )^2}  \\ \nonumber
              &=& \int\frac{d^2 {\bf p}}{(2\pi)^2} \int \frac{d \omega}{2\pi}
              \frac{m + 2 - {\rm cos}\,p_1 -{\rm cos}\,p_2}{\omega^2  + H({\bf p} )^2} \sigma_3
\end{eqnarray}
does not depend on the space coordinate, and   $\Xi^{(0)}_0 $ can be labelled as $\xi\sigma_3$.
Quantity $\lambda \xi$ in the Green function $G^{(0)}_{W}(x,p)$ may be considered as the correction to the fermion mass. To obtain the value of $\xi$,
we compute numerically the integral of Eq.(\ref{Xi_00}). The dependence of $\xi$ on $m$ is shown on Fig. \ref{fig.1}.
If coupling constant $\lambda$ is sufficiently large,
the 4 - fermion interaction will change essentially the value of
the effective mass parameter $m - \lambda \xi$.
As a result the system drops into the phase with the value of ${\cal N}_3$ different from that of the model without interactions.
The expression for the electric current, which follows from the relation { $\delta {\rm log}\,Z=J^k(x)\,\delta A_k(x)$}, gives
the Hall current
\rv{\begin{eqnarray}\label{current_2D}
J^k(x)& = & - \int \frac{d^3 p}{(2\pi)^3}  Tr G_{\lambda,W}(x,p) \frac{\partial}{\partial p_k}[G^{(0)}_{0,W}(x,p)]^{-1}\nonumber\\
      & = & - \int \frac{d^3 p}{(2\pi)^3}  Tr G_{\lambda,W}(x,p) \frac{\partial}{\partial p_k}[G^{(0)}_{\lambda,W}(x,p)]^{-1}
\end{eqnarray}}
Again, solving iteratively the Groenewold equation for $G_{\lambda,W}$ one comes to
\begin{equation}\label{HALL_k}
{j}^k_{Hall} = \cor{-}\frac{1}{2\pi}\,{\cal N}\,\epsilon^{ki} E_i ,
\end{equation}
where the topological invariant ${\cal N}$ is to be calculated using
the Green function ${\cal G}_\lambda = [i\omega  - H({\bf p})-\lambda \Xi^{(0)}_0]^{-1} $ as follows
\begin{eqnarray}\label{N3Ai_a}
{\cal N} &=&  -\frac{1}{24 \pi^2} {\rm Tr}\, \int_{} {\cal G}_\lambda^{-1} d {\cal G}_\lambda \wedge d {\cal G}_\lambda^{-1} \wedge d {\cal G}_\lambda
\end{eqnarray}

\subsection{Yukawa interaction}
\label{SectScalarA}

In this section, we consider the $2+1$ D tight-binding model with
interactions caused by scalar excitations. Let us start by consideration of Yukawa interactions, but our conclusions remain valid for the exchange by the wide class of excitations (including the most relevant case of Coulomb interactions to be discussed further). The Euclidean action is
\begin{eqnarray}\label{action_Yukawa3D}
S_{\eta}=S_0
+ \int d\tau \sum_{{\bf x,x'}}\phi_{\bf x'} \Big( \partial^2_{\tau}\delta_{\bf x,x'}+ {\cal B}_{\bf x',x}  \Big) \phi_{\bf x}
-\eta \sum_{\bf x}\bar{\psi}(\tau,{\bf x})\psi(\tau, {\bf x})\phi(\tau, {\bf x}).
\end{eqnarray}
where matrix
\begin {eqnarray}\label{bosonic_lattice_diff}
{\cal B}_{\bf x',x}=\sum_{i=1,2}(\delta_{x',x+e_i}+\delta_{x',x-e_i})\cor{-(M^2+4)}\delta_{x',x}
\end{eqnarray}
corresponds to the boson $\phi$ with mass $M$.  {Eq. (\ref{action_Yukawa3D}) includes Yukawa interaction,
which contributes to the self-energy of the fermions.
The leading order contribution is proportional to $\eta^2$. }
In the present work we consider its effect up to the order of $\eta^2$.

The electric current is given by
\begin {eqnarray}\label{current_3D_a}
J^k_\eta(x) =- \int \frac{d^3 p}{(2\pi)^3}  Tr G_{\eta,W}(x,p) \frac{\partial { Q_W    } }{\partial p_k}
\end{eqnarray}
where $G_{\eta,W}(x,p)=G^{(0)}_{\eta,W}(x,p)+G^{(1)}_{\eta,W}(x,p)+...$ is the full Green function with the interactions taken into account. The term $G^{(k)}_{\eta,W}(x,p)$ is proportional to the product of $k$ derivatives
$\frac{\partial}{\partial x}$. In particular,
\begin {eqnarray}\label{Green_W_express_Yukawa3D}
G^{(0)}_{\eta,W}(x,p) =[i(\omega - A_3(x)) - H({\bf p} - {A}(x))-\eta^2 \Sigma (p,x)  ]^{-1},
\end{eqnarray}
where the self-energy function $\Sigma=\eta^2 \Sigma_1+\eta^4 \Sigma_2+...$.
At the leading order, $\Sigma_1(x,p)$ is given by
\begin{eqnarray}\label{self-energy_a}
\Sigma_1(x,p) & = & - \int G_{0,W}(x,q) D(p-q) \frac{d^3 q}{(2\pi)^3} \nonumber\\
& =&\Sigma_1^{(0)}+\Sigma_1^{(1)}+...,
\end{eqnarray}
which is also expanded in powers of $\frac{\partial}{\partial x}$, according to the expansion of $G_{0,W}$.
More presisely, the Green function $G^{(0)}_{\eta,W}$ in
Eq.(\ref{Green_W_express_Yukawa3D})  should be written as
\begin {eqnarray}\label{Green_W_express_Yukawa3D_a}
G^{(0)}_{\eta,W}(x,p) =[i(\omega - A_3(x)) - H({\bf p} - {A}(x))-\eta^2 \Sigma^{(0)} (p,x)  ]^{-1}.
\end{eqnarray}

The  bosonic Green function is
\begin {eqnarray}\label{bosonic_Green}
D(p) = \frac{1}{\omega^2 + {\rm sin}^2 \cor{(p_1/2)} + {\rm sin}^2 \cor{(p_2/2)} + M^2}.
\end{eqnarray}
Contrary to the leading-order contribution of the 4 - fermion
interaction (which is  constant), the contribution of Yukawa interactions to the self-energy depends both on momenta and space coordinates.
Let us consider the difference $\mathcal{K}=J^k_{\eta}-J^k_{\eta \eta}$, where
$$
J_{\eta\eta} = \rv{-\int d^3 p}\, {\rm Tr}\, G^{}_{\eta,W}(x,p) \frac{\partial}{\partial p_k}({ Q_W - \Sigma_W    })
$$
Then
\begin {eqnarray}\label{current_change}
\mathcal{K}
&=& - \int \frac{d^3 p}{(2\pi)^3} \, {\rm Tr}\, G_{\eta,W}(x,p) \frac{\partial}{\partial p_k} \Sigma(x,p)  \nonumber \\
&=&  -\eta^2\int \frac{d^3 p}{(2\pi)^3} \, {\rm Tr}\, G_{0,W}(x,p) \frac{\partial}{\partial p_k} \Sigma_1(x,p) + O(\eta^4).
\end{eqnarray}
Using  expansion  $G_{0,W}= G^{(0)}_{0,W}(x,p)+ G^{(1)}_{0,W}(x,p)+...$
with $G^{(n)}_{\eta,W} \sim O(\partial_x^n)$,
we can represent correspondingly: $\mathcal{K}= \mathcal{K}^{(0)}+ \mathcal{K}^{(1)}+...$ .
Let us first consider $ \mathcal{K}^{(0)}$ and denote for simplicity $G^{(0)}_{0,W}(x,p)$ by $g(p)$:
\begin {eqnarray}\label{current_change_LO}  
\mathcal{K}^{(0)}
&=& \eta^2\int \frac{d^3 p}{(2\pi)^3} Tr \big[ G^{(0)}_{0,W}(x,p)
        \frac{\partial}{\partial p_k} \int G^{(0)}_{0,W}(x,p-q)D(q)\frac{d^3 q}{(2\pi)^3}\big]  \nonumber \\
&=& \eta^2\int \frac{d^3 p}{(2\pi)^3} Tr \big[ g(p)
\frac{\partial}{\partial p_k} \int g(p-q)D(q)\frac{d^3 q}{(2\pi)^3} \big], \nonumber
\end{eqnarray}
in which higher order corrections $O(\eta^4)$ have been omitted.
If we denote $\mathcal{K}^{(0)}= \eta^2 I$, then
\rv{\begin {eqnarray}\label{current_change_LO_2}  
I &=&  \int \frac{d^3 p}{(2\pi)^3}\frac{d^3 q}{(2\pi)^3}
{\rm Tr} \big[ g(p)\frac{\partial}{\partial p_k} g(p-q) \big] D(q)   \nonumber \\
&=& -\int \frac{d^3 p}{(2\pi)^3}\frac{d^3 q}{(2\pi)^3}
{\rm Tr} \big[ \frac{\partial g(p)}{\partial p_k}  g(p-q)\big] D(q)   \nonumber \\
&=& -\int \frac{d^3 p}{(2\pi)^3}\frac{d^3 q}{(2\pi)^3}
{\rm Tr} \big[ g(p-q) \frac{\partial g(p)}{\partial p_k}  \big] D(q)   \nonumber \\
&=& -\int \frac{d^3 s}{(2\pi)^3}\frac{d^3 q}{(2\pi)^3}
{\rm Tr} \big[ g(s) \frac{\partial g(s+q)}{\partial s_k}  \big] D(q)   \nonumber \\
&=& -\int \frac{d^3 s}{(2\pi)^3}\frac{d^3 t}{(2\pi)^3}
{\rm Tr} \big[ g(s) \frac{\partial g(s-t)}{\partial s_k}  \big] D(-t).   \nonumber
\end{eqnarray}}
Since $D(-t)=D(t)$, one finds that $I=-I$, therefore $I=0$, which implies
$\mathcal{K}^{(0)} =0$.

Now, we consider the next order in the derivatives of the gauge field.
\begin {eqnarray}\label{current_change_NLO}  
 \mathcal{K}^{(1)}& =& \eta^2\int \frac{d^3 p}{(2\pi)^3}
\, \Big({\rm Tr}\,  G^{(1)}_{0,W}(x,p)\frac{\partial }{\partial p_k} \Sigma^{(0)}(x,p)
\\&&+  {\rm Tr}\,  G_{0,W}^{(0)}(x,p)\frac{\partial }{\partial p_k} \Sigma^{(1)}(x,p)\Big)   \nonumber \\
&=&\eta^2 \int_p \int_q
\, \Big({\rm Tr}\,  G_{0,W}^{(1)}(x,p) G_{0,W}^{(0)}(x,q)  \frac{\partial}{\partial p_k} D(p-q)
\nonumber\\&&+  {\rm Tr}\,  G_{0,W}^{(0)}(x,p)G_{0,W}^{(1)}(x,q)\frac{\partial }{\partial p_k} D(p-q)   \Big)\nonumber \\
&=&\eta^2\int_p \int_q
\, \Big({\rm Tr}\,  G_{0,W}^{(1)}(x,p) G_{0,W}^{(0)}(x,q) \frac{\partial}{\partial p_k} D(p-q)
\nonumber\\&&+  {\rm Tr}\,  G_{0,W}^{(1)}(x,q)G_{0,W}^{(0)}(x,p)\frac{\partial }{\partial p_k} D(p-q)  \Big) \nonumber \\
&=&\eta^2\int_p \int_q
\, \Big({\rm Tr}\,  G_{0,W}^{(1)}(x,q) G_{0,W}^{(0)}(x,p)  \frac{\partial}{\partial q_k} D(q-p)
\nonumber\\&&+  {\rm Tr}\,  G_{0,W}^{(1)}(x,q)G_{0,W}^{(0)}(x,p)\frac{\partial }{\partial p_k} D(p-q)  \Big) \nonumber \\
&=& 0
\end{eqnarray}
Here, we used that $D(-t) = D(t)$, and $D'(-t)=-D'(t)$. Therefore, we obtain $\delta J^{k,(1)} =\mathcal{K}^{(1)}=0$ and conclude that, at least in the one - loop appoximation, Yukawa interactions do not affect the expression for the Hall conductivity in terms of the (interacting) Green function. As it was mentioned above, in the same way it may be proved that the Hall conductivity is not affected (up to the term $\sim \eta^2$) by the exchange by scalar boson with arbitrary propagator $D(p)$ obeying $D(p) = D(-p)$. The important particular case is when $D(p)$ is the (three - dimensional) Coulomb interaction. It is resulted from the interactions due to the exchange by real photons between the Bloch electrons of the given $2D$ material.
The Hall current is given by
\begin{equation}\label{HALL_k_b}
{j}^k_{Hall} =\cor{-} \frac{1}{2\pi}\,{\cal N}\,\epsilon^{ki} E_i ,
\end{equation}
where the topological invariant ${\cal N}$ is to be calculated using
the interacting Green function
${\cal G}_\eta = [i\omega  - H({\bf p})-\eta^2 \Sigma_1^{(0)}(p)]^{-1} $
(with $\Sigma_1^{(0)}(p) = -\int {\cal G}_{0}(q) D(p-q) d^3 q/(2\pi)^3$):
\begin{eqnarray}\label{N3Ai_b}
{\cal N} &=& \rv{- \frac{1}{24 \pi^2}} {\rm Tr}\, \int_{} {\cal G}_\eta^{-1} d {\cal G}_\eta \wedge d {\cal G}_\eta^{-1} \wedge d {\cal G}_\eta
\end{eqnarray}
The extension of this result to the three - dimensional materials is also straightforward and will be considered in the next section.

\subsection{Coulomb interaction and Topological phase transitions}

In this subsection, we consider the $2+1$ D tight-binding model with
Coulomb interaction. The Euclidean action is
\begin{eqnarray}\label{action_Coulomb3D}
S= S_0-\alpha \int d\tau \sum_{{\bf x,x'}}
\bar{\psi}(\tau,{\bf x})\psi(\tau, {\bf x})V({\bf x-x'})\bar{\psi}(\tau,{\bf x'})\psi(\tau, {\bf x'}),
\end{eqnarray}
where $V$ is the Coulomb potential $V({\bf x})=1/|{\bf x}|=1/\sqrt{x_1^2+x_2^2}$.
There is no magnetic field in this systems while there is the external electrical field.
We chose the gauge with vanishing $A_1$ and $A_2$.
The electric current is given by Eq. (\ref{HALL_k_b}) with $\cal N$ given by Eq. (\ref{N3Ai_b}).
Here, the Green function entering this expression is given by
{
\begin {eqnarray}\label{Green_W_express_Coulomb3D}
{\cal G}_{\alpha}(p) =[i\omega  - H({\bf p} )-\alpha \Sigma(p)  ]^{-1}+O(\alpha^2) \nonumber
\end{eqnarray}    } 
with
\rev{\begin{eqnarray}\label{self-energy_b}
\Sigma(p) & = & - \int {\cal G}_{\alpha = 0}(q) \tilde{V}(p-q) \frac{d^3 q}{(2\pi)^3}
\end{eqnarray}}
$\tilde{V}(p)$ is the Coulomb potential in the momentum space
$\tilde{V}(p) =\sum_{\bf x} e^{i{\bf p\cdot x}}/ \sqrt{x_1^2+x_2^2}$.
Therefore, $\Sigma(p)$ depends only on $p_1$ and $p_2$.
Similar to the results of the previous subsections, Coulomb interaction does not
change the Hall conductivity until the topological phase transition is encountered, at least to the
leading order in $\alpha$. Without interactions the topological phase transitions occur
at $m = 0, -2, -4$. The self-energy function contribution to the Green function modifies those critical values.
Therefore, the values of $m$ that give rise to a certain value of ${\cal N}$
without interactions may lead to the different value of $\cal N$ in the presence of interactions.

In order to investigate the effect of Coulomb interactions on the critical values $m^\prime, m^{\prime \prime}, m^{\prime \prime \prime}$ of $m$
we consider the effect of $\Sigma$ on the poles of the Green's function.
Let us denote ${\cal G}_\alpha^{-1}(p)=i\sigma^3(\sigma^k g_k(p)-ig_4(p))$ with $k=1,2,3$.
Recall that \cite{Z2016_1} (see also Appendix B)
\rv{\begin{eqnarray}
{\cal N} &=&   - \frac{1}{2}\sum_l \, {\rm sign}(g_4(y^{(l)})) \,{\rm Res}\,(y^{(l)})
\end{eqnarray}   }
where
\begin{eqnarray}
{\rm Res}\,(y) &=&  \frac{1}{8 \pi} \epsilon^{ijk}\, \int_{\partial \Omega(y)} \, v_i  d v_j \wedge d v_k
\end{eqnarray}
and  $v_k=g_k/\sqrt{g_1^2+g_2^2+g_3^2}$ with $k=1,2,3$.
By $y^{(l)}$ we denote the positions of the zeros of function $g_1^2+g_2^2+g_3^2$.

It may be shown that in the first order in $\alpha$, the Coulomb interactions
do not change the positions of $y^{(l)}$, that are
$$
y^{(1)} = (0,0,0),\quad y^{(2)} = (0,0,\pi),$$
$$ \quad y^{(3)} = (0,\pi,0),\quad y^{(4)} = (0,\pi,\pi)
$$
Let us denote \rev{$\Sigma(p)=-i\sigma^3(\sigma^k f_k(p)-if_4(p))$}. It is easy to find that
$f_k(y^{(m)})=0$ for $k=1,2,3$ because $\tilde{V}(p-y^{(l)}) = \tilde{V}(-p-y^{(l)})$.
Therefore, the self energy affects the Green function at $p = y^{(l)}$ only through the modification of $g_4$ by $f_4$,
which is given by
\begin{eqnarray}
&&f_4(p_1,p_2|m)=\nonumber\\&&\int \frac{m + 2-{\rm cos}\,q_1-{\rm cos}\,q_2}
{\sqrt{{\rm sin}^2\,q_1 + {\rm sin}^2\, q_2 + (m + 2-{\rm cos}\,q_1-{\rm cos}\,q_2)^2}}
\nonumber\\&&\tilde{V}(p_1 - q_1,p_2 - q_2) \frac{d^2 q}{2(2\pi)^2}.
\end{eqnarray}
The critical values of $m$ appear as the solutions of equations:
$$
m^\prime+\alpha f_4(0,0|m^\prime)=0,$$$$ \, m^{\prime\prime}+2+\alpha f_4(0,\pi|m^{\prime\prime})=0,$$$$\, m^{\prime\prime\prime}+4+\alpha f_4(\pi,\pi|m^{\prime\prime\prime})=0
$$

To the leading order in $\alpha$, these critical values are given by
$m^\prime=-\alpha f_4(0,0|m=0)$, $m^{\prime\prime}=-2-\alpha f_4(0,\pi|m=-2)$
and $m^{\prime\prime\prime}=-4-\alpha f_4(\pi,\pi|m=-4)$. We found analytically
that $f_4(\pi,\pi|m=-4)=-f_4(0,0|m=0)$ and $f_4(0,\pi|m=-2)=0$. As for
the numerical value of $f_4(0,0|m=0)$, we computed the related integral on the $40\times 40$ lattice,
and the numerical result is -0.28.

\begin{figure}[h]
	\centering  %
	\includegraphics[width=0.5\linewidth]{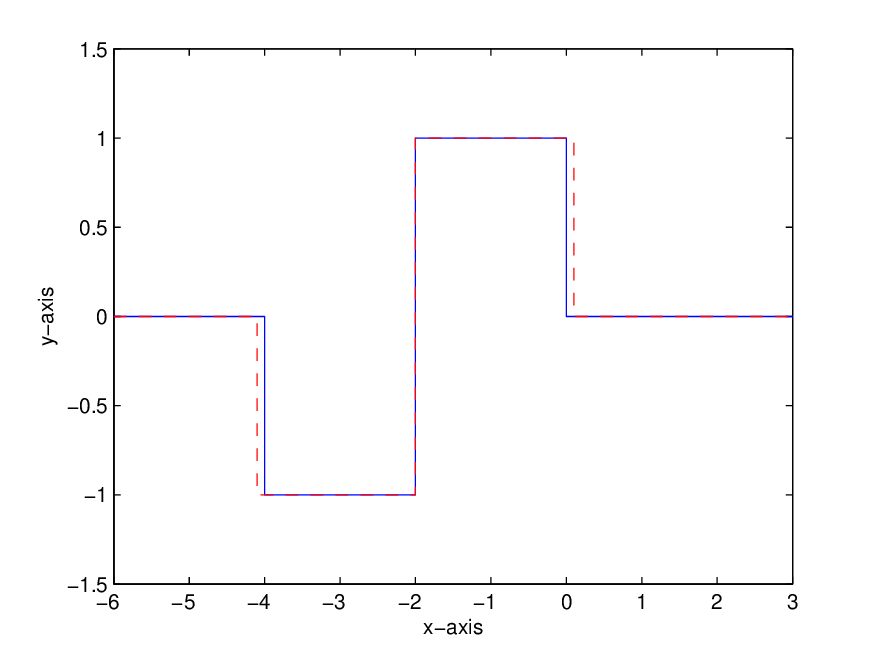}  %
	\caption{The dependence of ${\cal N}$ (y-axis) on mass parameter $m$ (x-axis).
             The blue solid line represents ${\cal N}(m)$ without interaction, while the
             red dashed line is ${\cal N}(m)$ with Coulomb interactions taken into account to the first
             order in $\alpha$, with $\alpha=0.3$.}  %
	\label{fig.3}   %
\end{figure}
Up to the leading order in $\alpha$, the topological number ${\cal N}$ is represented in Fig.\ref{fig.3} as a function of $m$ for $\alpha = 0.3$.


\subsection{Higher-order corrections}

In this subsection, we extend the above consideration to the calculation of radiative corrections to Hall conductivity. To some extent the results obtained below repeat those of  our previous work \cite{Zhang_2019_JETPL}.
We will consider the AQHE in the considered above system without magnetic field, and the interactions between the fermions due to exchange by scalar bosons, { shown in Eq.(\ref{action_Yukawa3D}).}

{ In Sec.2.3 above,} we considered effects of interactions to the order $O(\eta^2)$.
In this subsection, we extend our result to the higher orders, with the help of diagrammatics.

\begin{figure}[h]
	\centering  %
	\includegraphics[width=0.3\linewidth]{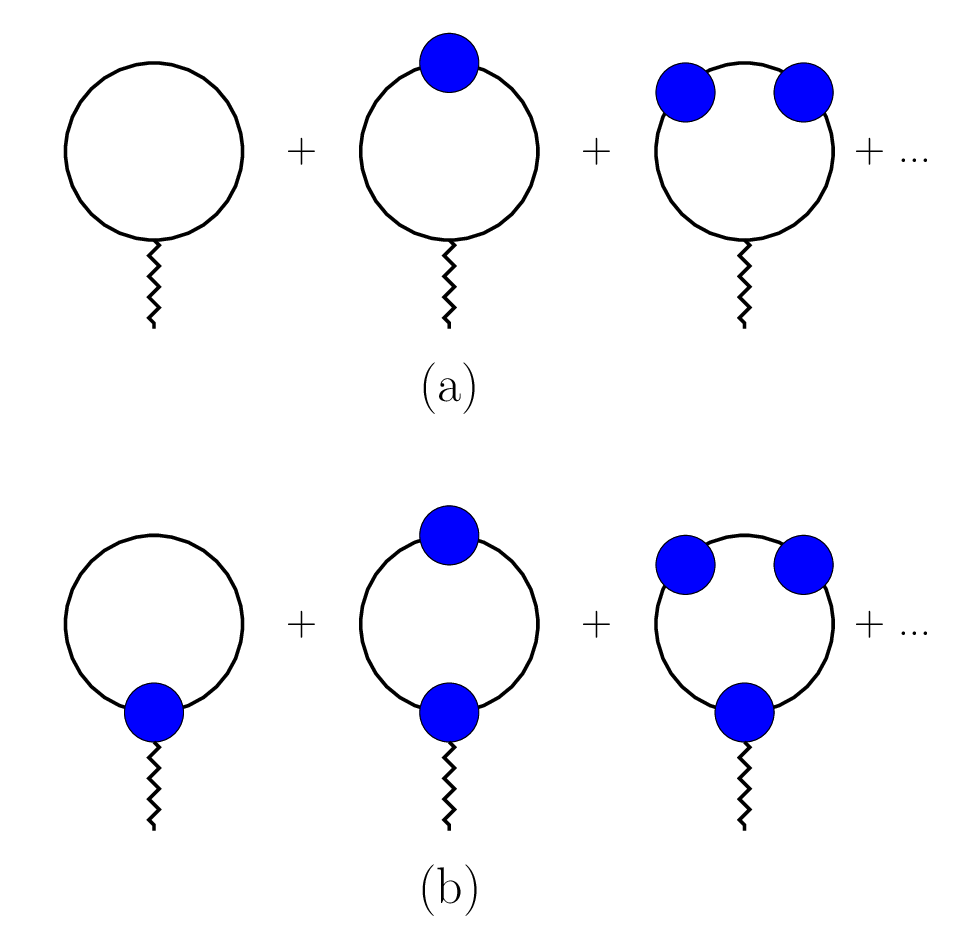}  %
	\caption{Tadpole diagrams. The black solid lines are the propagators of fermions,
while the black zigzag represents an external field. The shaded blue circles correspond to the self-energy functions.
{ (a) Diagrams related to the current density $J^k_{\eta}$.
(b)Diagrams related to the difference $J^k_{\eta} - J^k_{\eta\eta}$.    }   }  %
	\label{fig_tadpole_general}   %
\end{figure}

Here we need an assumption that $G_W(x,p)$ and $\Sigma_W(x,p)$ are
functions of $p-A(x)$, i.e. their dependence on $x$ and $p$ only
comes from the combination $p-A(x)$. This can be shown easily order by order in $\eta$.
Taking advantage of this, it is found that the current density is
\begin {eqnarray}\label{current_star}
J_k(x) &=& \cor{-}\int \frac{d^3 p}{(2\pi)^3} \, Tr G_W(x,p) \partial_{p_k} Q \nonumber \\
      &=& \cor{-}\int \frac{d^3 p}{(2\pi)^3} \, Tr G_W(x,p)\star \partial_{p_k} Q
\end{eqnarray}
where we changed the ordinary product to star product because of the above mentioned assumption and also because we intend to consider response to constant field strength.
Therefore, the current density can be written as
$J^k_{\eta} =\cor{-} \int \frac{d^3 p}{(2\pi)^3} \,Tr G_{\eta} \star \partial_{p_k} Q$,
while $\mathcal{K}=J^k_{\eta}-J^k_{\eta \eta}= \cor{-}\int \frac{d^3 p}{(2\pi)^3}  \, Tr G_{\eta} \star \partial_{p_k} \Sigma$.
Using the diagram technique proposed in \cite{ZZ2019_3} we expand $J^k_{\eta}=\cor{-}\sum_{n=0}^{\infty} \mathcal{J}[n]$, where
\begin {eqnarray}\label{J_component}
\mathcal{J}[n] = \cor{-}\int \frac{d^3 p}{(2\pi)^3} \, Tr (G_{0}\star \Sigma \star)^n G_0  \star \partial_{p_k} Q,
\end{eqnarray}
This expansion is represented in Fig.\ref{fig_tadpole_general}(a).
Similarly,  $\mathcal{K}$ is shown in Fig.\ref{fig_tadpole_general}(b),
and can be expressed as $\mathcal{K}=\sum_{n=0}^{\infty} \mathcal{K}[n]$, with
\begin {eqnarray}\label{deltaJ_component}
\mathcal{K}[n] =\cor{-} \int \frac{d^3 p}{(2\pi)^3} \,  Tr (\Sigma \star G_0\star)^n G_{0} \star  \partial_{p_k} \Sigma.
\end{eqnarray}
In our previous work \cite{Zhang_2019_JETPL} we found relation between $J^k_{\eta} $ and  $\mathcal{K}$,
and have shown (using integration by parts and some algebra) that for $n\geq 1$,
\begin {eqnarray}\label{deltaJ_and_J}
\mathcal{K}[n] = \mathcal{J}[n+1] .
\end{eqnarray}

\begin{figure}[h]
	\centering  %
	\includegraphics[width=0.3\linewidth]{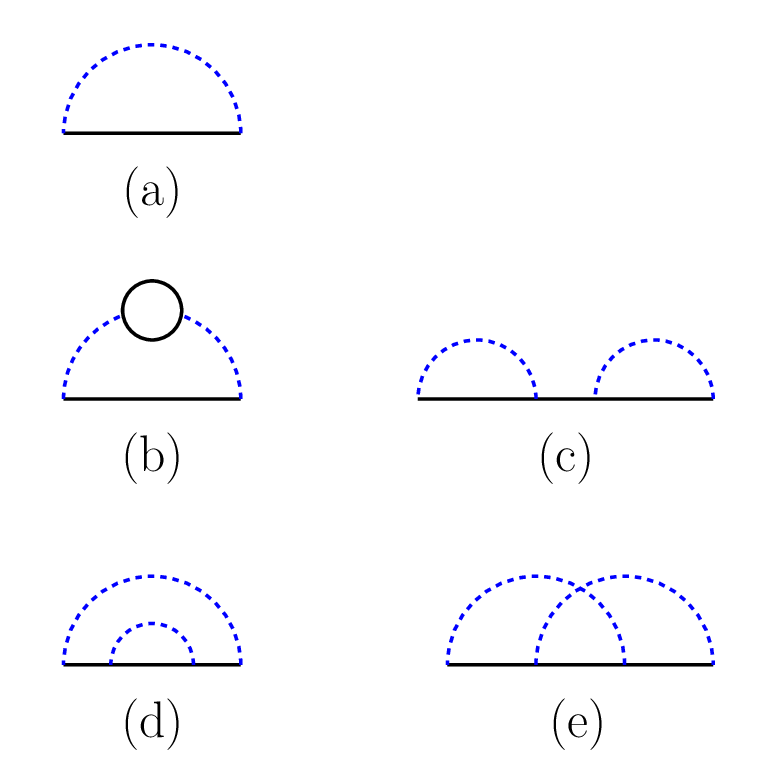}  %
	\caption{Self-energy functions. The dashed blue lines correspond to
the bosons responsible for the Yukawa interaction.}  %
	\label{fig_self-energy}   %
\end{figure}

\begin{figure}[h]
	\centering  %
	\includegraphics[width=0.3\linewidth]{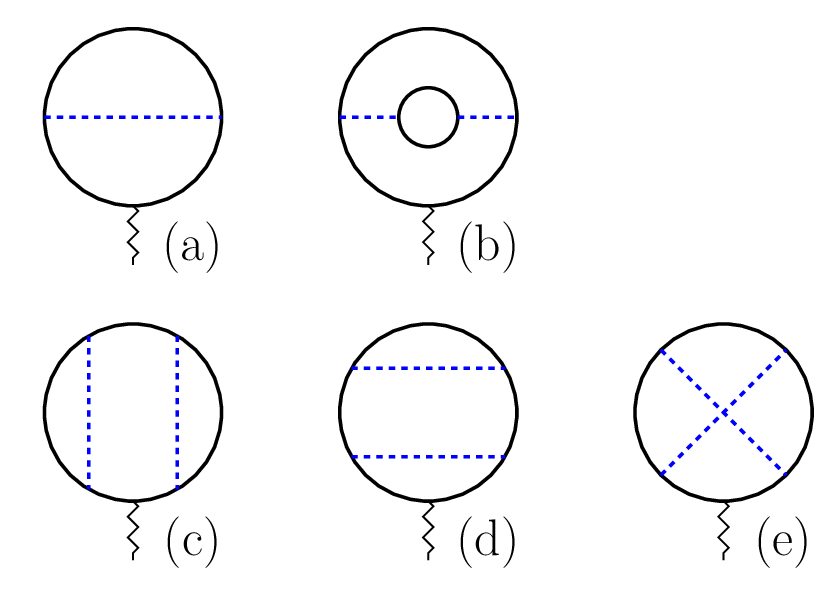}  %
	\caption{Tadpole graphs in the first and the second order.}  %
	\label{fig_tadpole}   %
\end{figure}

\begin{figure}[h]
	\centering  %
	\includegraphics[width=0.2\linewidth]{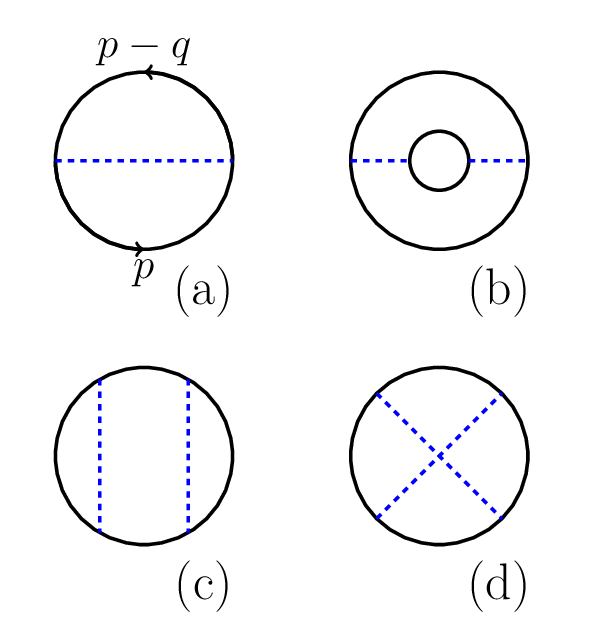}  %
	\caption{Graphs of bubble-like "progenitors" in the first and the second order.}  %
	\label{fig_bubbles}   %
\end{figure}

\begin{figure}[h]
	\centering  %
	\includegraphics[width=0.2\linewidth]{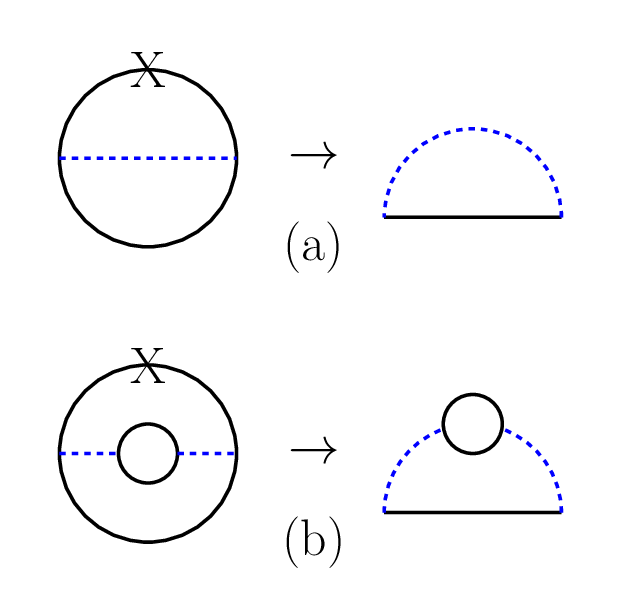}  %
	\caption{"Progenitors" and the corresponding self-energies, in the first and the second order.}  %
	\label{fig_cut-glue_1}   %
\end{figure}

In what follows, we will prove that $\mathcal{K}=0$
through the diagrammatic approach,  to order $\eta^4$. The consideration of the higher orders is completely similar.
Let us  consider $\mathcal{K}_1$ first (this corresponds to
the interaction contribution to $\mathcal{K}$ proportional to $\eta^2$).
\begin {eqnarray}\label{K1(0)}
\mathcal{K}_1&=& \cor{+}\int \frac{d^3 p}{(2\pi)^3} {\rm Tr}  \Sigma_{1,W} (R,p) \star \partial_{p_k} G_{0,W}(R,p) \\
&=& -\int \frac{d^3 p}{(2\pi)^3}\int \frac{d^3 q}{(2\pi)^3} {\rm Tr}  G_{0,W}(R,p-q)D(q) \star \partial_{p_k} G_{0,W}(R,p) ,
\end{eqnarray}
where $\Sigma_{1,W}$ is shown by Fig.\ref{fig_self-energy}(a).
$\mathcal{K}_1$ as a whole can be shown by Fig.\ref{fig_tadpole}(a).
Notice that the last expression without $\partial_{p_k}$
corresponds to the bubble diagram shown in Fig.\ref{fig_bubbles}(a),
which is expressed as
\begin {eqnarray}\label{bubble_1}
\mathcal{B}_1
=-\int \frac{d^3 p}{(2\pi)^3}\int \frac{d^3 q}{(2\pi)^3} {\rm Tr}  [G_{0,W}(R,p-q) \star G_{0,W}(R,p)] D(q),
\end{eqnarray}
and is called "progenitor" (see \cite{Zhang_2019_JETPL} and references therein).
We can understand the meaning of the term "progenitor" in the following way.
Because of the integration over $p$, inserting a partial derivative $\partial_{p_k}$ to the
integrand of $\mathcal{B}_1$ gives zero:
\begin {eqnarray}\label{bubble_1_partial}
\int \frac{d^3 p}{(2\pi)^3}\int \frac{d^3 q}{(2\pi)^3}
   {\rm Tr}  \partial_{p_k} [G_{0,W}(R,p-q) \star G_{0,W}(R,p)]D(q)=0
\end{eqnarray}
Operation $\partial_{p_k}$ produces two terms (the product rule in differential calculus),
and it can be shown through integration by parts that
they have equal values
(also shown above in subsection \ref{SectScalarA}).
Each term is equal to
the Feynman diagram shown in Fig.\ref{fig_tadpole} (a).
Therefore,
adding the derivative $\partial_{p_k}$ to the expression of $\mathcal{B}_1$,
produces or "generates" the expression of $\mathcal{K}_1$.
Alternatively, we can understand
what is "progenitor" in a diagrammatic way: cutting and
erasing the fermion line marked by cross "X" in Fig.\ref{fig_cut-glue_1} (a), we
obtain the self-energy shown in  Fig.\ref{fig_self-energy}(a). Therefore,
The cut/glue action connects a bubble-like "progenitor"
to the self-energy diagram.

\begin{figure}[h]
	\centering  %
	\includegraphics[width=0.2\linewidth]{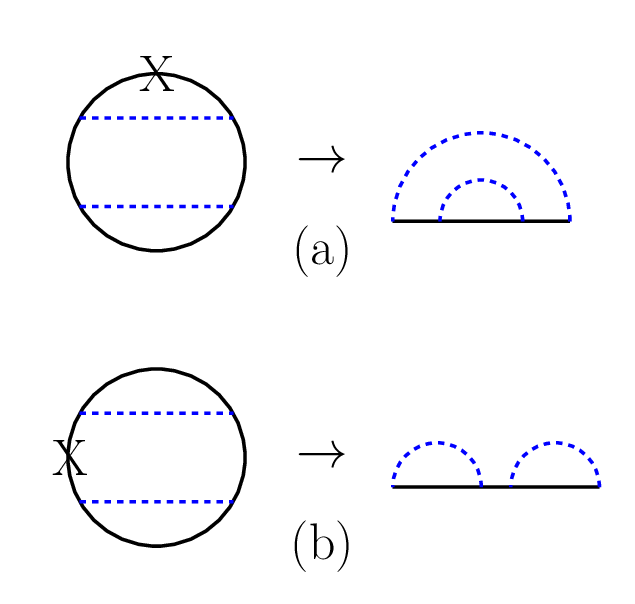}  %
	\caption{"Progenitors" and the corresponding self-energies, in the second order (non-entangled case).}  %
	\label{fig_cut-glue_2}   %
\end{figure}

\begin{figure}[h]
	\centering  %
	\includegraphics[width=0.2\linewidth]{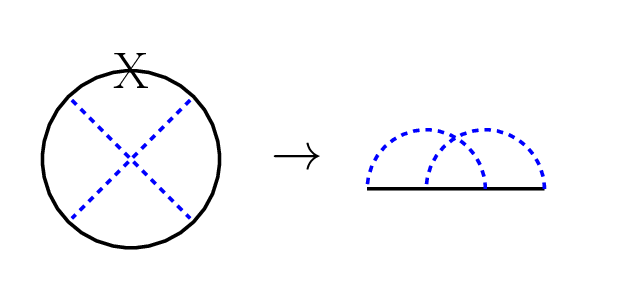}  %
	\caption{"Progenitor" and the corresponding self-energy, in the second order (the entangled case).}  %
	\label{fig_cut-glue_3}   %
\end{figure}

From Eq.(\ref{bubble_1_partial}), we know that  $\mathcal{K}_1=0$.
Furthermore, from Eq.(\ref{deltaJ_and_J}), we obtain $\mathcal{J}_1=0$,
which means the interaction contribution to the current density vanishes at
the leading order in the coupling $O(\eta^2)$.
Our next step is to analyze $\mathcal{K}_2$ (related to the order $O(\eta^4)$):
\begin {eqnarray}\label{K2(0)}
\mathcal{K}_2 = \cor{+}\int \frac{d^3 p}{(2\pi)^3} {\rm Tr} \Xi_{2,W} (R,p) \star \partial_{p_k} G_{0,W}(R,p)
\end{eqnarray}
in which $\Xi_{2,W} = \Sigma_{2,W} + \Sigma_{1,W} \star G_{0,W} \star \Sigma_{1,W}$
is shown by the diagrams of Fig.\ref{fig_self-energy}(b-e),
and the corresponding diagrams for $\mathcal{K}_2$ are shown in
Fig.\ref{fig_tadpole}(b-e).
Among them, the case of Fig.\ref{fig_tadpole}(b)
is relatively simple:
Similar to Fig.\ref{fig_tadpole}(a),
the diagram in Fig.\ref{fig_tadpole}(b) is also zero. The proof
is similar, and the only necessary change is to replace
$D(q)$ in Eq. (\ref{K1(0)}) into $D(q)\Pi(q^2)D(q)$, where $\Pi(q^2)$ is the vacuum
polarization function.
Alternatively, we can get the same result in the diagrammatic
way through the progenitor shown in Fig.\ref{fig_bubbles}(b).
Using Fig.\ref{fig_cut-glue_1}(b) we observe that
adding the crosses "X" to this progenitor produces the diagrams with the same pattern.
Therefore, the contribution of the self-energy in Fig.\ref{fig_self-energy}(b)
to the current is zero, i.e. Fig.\ref{fig_tadpole}(b) is zero.

Contribution of the self-energy shown in Fig.\ref{fig_self-energy}(c) and (d)
(rainbow diagrams, $r.b.$ for short ) can be evaluated as
\begin {eqnarray}\label{K2(0)_rainbow}
\mathcal{K}_{2,r.b.}
&=& \cor{+}\int \frac{d^3 p}{(2\pi)^3}  \int \frac{d^3 q}{(2\pi)^3}  \int \frac{d^3 k}{(2\pi)^3}
 {\rm Tr}  [ G_{W}(R,p-q)D(q) \star G_{W}(R,p) \nonumber \\
 &&  \star G_{W}(R,p-k)D(k) \star \partial_{p_k}G_{W}(R,p) \nonumber \\
&& + G_{W}(R,p-q)D(q) \star G_{W}(R,p-q-k)D(k)\nonumber \\
&& { \star G_{W}(R,p-q)} \star \partial_{p_k}G_{W}(R,p)  ], \nonumber \\
\end{eqnarray}
which corresponds to Fig.\ref{fig_tadpole}(c) and (d).
From the bubble-like progenitor in Fig.\ref{fig_bubbles}(c),
adding the crosses, we obtain two different patterns shown in
Fig.\ref{fig_cut-glue_2}. Therefore, the total contribution of
the self-energy functions Fig.\ref{fig_self-energy}(c) and (d)
is zero, i.e. the sum of the diagrams Fig.\ref{fig_tadpole}(c) and (d)
is zero. Finally, the contribution of the cross diagram in Fig.\ref{fig_self-energy}(e),
(also in Fig.\ref{fig_tadpole}(e) ) is even simpler:
adding the crosses to the bubble in Fig.\ref{fig_bubbles} (d)
produces 4 diagrams with the same pattern shown in
Fig.\ref{fig_cut-glue_3}.
Therefore, the diagram Fig.\ref{fig_tadpole}(e) is zero.
Up to now, we proved $\mathcal{K}_1=\mathcal{K}_2=0$.
Then, from Eq.(\ref{deltaJ_and_J}), we obtain $\mathcal{J}_1=\mathcal{J}_2=0$,
which means the interaction contribution to the current density vanishes at
the orders $O(\eta^2)$ and $O(\eta^4)$. Consideration of the higher orders is similar.




\section{AQHE in the $3+1$ D Weyl semimetal with interactions}
\label{Sec_4D}

\subsection{$3+1$ D Weyl semimetal with four-Fermi interaction}

In this section, we consider the particular model of Weyl semimetals with the four - fermion interaction  in $3+1$ D space-time.
The Euclidean action is
\begin{eqnarray}\label{action_4D}
S_{\lambda}& =& \int d\tau \sum_{{\bf x}}\Big[\bar{\psi}\Big(i(i \partial_{\tau} - A_4(-i\tau,{\bf x})) \\&&  - H(-i\partial_{\bf x} - {\bf A}({\bf x}))\Big)\psi
+\frac{\lambda}{2} (\bar{\psi}(\tau,{\bf x})\psi(\tau, {\bf x}))^2\Big],\nonumber
\end{eqnarray}
where
\begin{equation}\label{Ham_4D}
H({\bf p}) = {\rm sin}\,p_1\, \sigma^2 - {\rm sin}\, p_2 \, \sigma^1 - (m - {\rm cos}\,p_3
             + \sum_{i=1,2}(1-{\rm cos}\,p_i)) \, \sigma^3
\end{equation}
with $m \in (-1,1)$. This system contains the two Fermi points
\begin{equation}\label{Fermi_point0}
{\bf K}_{\mp}=(0,0,\pm \beta,0),\quad \beta={\rm arccos} \, m.
\end{equation}
Here ${\bf K}_+$ is the right - handed Weyl point while ${\bf K}_-$ is the left - handed one.
As in the $2+1$ D model, the electric current in the $3+1$ D Weyl semimetal
is given by
\rv{\begin{eqnarray}\label{current_4D}
J^k(x)& = & - \int \frac{d^4 p}{(2\pi)^4}  Tr G_{\lambda,W}(x,p) \frac{\partial}{\partial p_k}[G^{(0)}_{0,W}(x,p)]^{-1}\nonumber\\
&=&J^{(0),k}+J^{(1),k}+...
\end{eqnarray}}

\begin{figure}[h]
	\centering  %
	\includegraphics[width=0.5\linewidth]{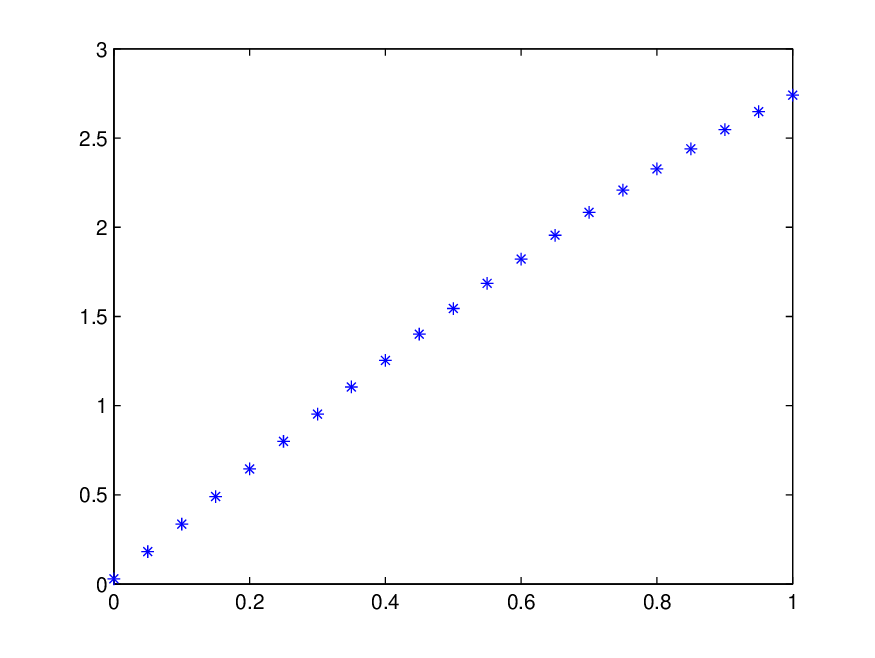}  %
	\caption{The dependence of $\kappa$ {(related to $K^{(0)}_0$ in Eq.(\ref{H'_4D}))}
              on mass parameter $m$.
             The horizontal axis represents the dimensionless mass parameter,
             i.e. the physical value of $m$ multiplied by the lattice spacing.}  %
	\label{fig.2}   %
\end{figure}

In the leading order
\begin{eqnarray}\label{current_J0_4D}
J^{(0),k} &=& \rv{-\int \frac{d^4 p}{(2\pi)^4}}  Tr G^{(0)}_{\lambda,W}(x,p) \frac{\partial}{\partial p_k}[G^{(0)}_{0,W}(x,p)]^{-1}\nonumber\\& =& \rv{-\int \frac{d^4 p}{(2\pi)^4}}  Tr G^{(0)}_{\lambda,W}(x,p) \frac{\partial}{\partial p_k}[G^{(0)}_{\lambda,W}(x,p)]^{-1}\nonumber\\&
         =&0
\end{eqnarray}
where
\begin{eqnarray}\label{Green_W_express_4D}
G^{(0)}_{\lambda,W}(x,p) =[i(\omega - A_4(x)) - H({\bf p} - { A}(x))+\lambda K^{(0)}_0(x)]^{-1}   \nonumber  
\end{eqnarray}
with the corrections of the order of $\lambda^2$ neglected.
$K^{(0)}_0$ in the above equation is given by
\begin {eqnarray}\label{H'_4D}
 K^{(0)}_0  &=& -\int \frac{d^4 p}{(2\pi)^4}  \Big(G^{(0)}_{0,W}(x,p)-{\rm Tr} \, (G^{(0)}_{0,W}(x,p))\Big)   \nonumber \\
              &=&  -\int\frac{d^3 {\bf p} d \omega}{(2\pi)^4}
              \frac{m - {\rm cos}\,p_3 + 2 - {\rm cos}\,p_1 -{\rm cos}\,p_2}{\omega^2  + H({\bf p} )^2} \sigma_3,
\end{eqnarray}
which can be expressed by $\kappa \sigma_3$. $\kappa$
does not depend on the space coordinates and can be calculated
numerically (Fig. \ref{fig.2}). Comparing with Eq.(\ref{Ham_4D}),
we find that $\lambda \kappa$ is the correction to mass parameter.
As for the next to leading order, following \cite{Z2016_1} we obtain
\begin {eqnarray}\label{current_J1_4D}
J^{(1) k} &=&- \int \frac{d^4 p}{(2\pi)^4} \, {\rm  Tr}\, G^{(0)}_{\lambda,W}(x,p)
\frac{\partial}{\partial p_i}[G^{(0)}_{\lambda,W}(x,p)]^{-1}\nonumber\\&&
\frac{\partial}{\partial p_j}[G^{(0)}_{\lambda,W}(x,p)]{ F_{ij} }
\frac{\partial}{\partial p_k}[G^{(0)}_{0,W}(x,p)]^{-1}               \\ \nonumber
          &=&- \int \frac{d^4 p}{(2\pi)^4}  \, {\rm Tr}\,  G^{(0)}_{\lambda,W}(x,p)
\frac{\partial}{\partial p_i}[G^{(0)}_{\lambda,W}(x,p)]^{-1}\nonumber\\&&
\frac{\partial}{\partial p_j}[G^{(0)}_{\lambda,W}(x,p)]{ F_{ij} }
\frac{\partial}{\partial p_k}[G^{(0)}_{\lambda,W}(x,p)]^{-1}
\end{eqnarray}
Up to the terms linear in field strength $F_{ij}$, we obtain for the Hall current in $3+1$ D the result similar to that of \cite{Z2016_1}:
\begin{equation}\label{HALL_4D}
{j}^k_{Hall} = \cor{-}\frac{1}{4\pi^2}\,{\cal N}_l\,\epsilon^{kjl} E_j ,
\end{equation}
where
\begin{eqnarray}\label{M_4D}
{\cal N}_l &=& - \frac{1}{24 \pi^2}\epsilon^{ijkl}
 \int_{} d^4 p \, {\rm Tr}\, {\cal G}_\lambda  \frac{\partial {\cal G}_\lambda^{-1}}{\partial p_i}
                        \frac{\partial {\cal G}_\lambda}{\partial p_j} \frac{\partial {\cal G}_\lambda^{-1}}{\partial p_k}.
\end{eqnarray}

Without interactions (at $\lambda = 0$) the values of ${\cal N}_l$ were calculated in \cite{Z2016_1}. We repeat this calculation in
Appendix C for completeness. Notice that unlike the case of the insulators, for the Weyl semimetals the values of ${\cal N}_l$
are not topological invariants. However, unlike the case of ordinary metals with Fermi surfaces,
the corresponding integrals over momenta are convergent and the values of ${\cal N}_l$ are well - defined.

From \cite{Z2016_1} we know that
${\cal N}_1={\cal N}_2={\cal N}_4=0$, while ${\cal N}_3$ depends on mass parameter $m$.
When $m$ belongs to the interval $(-1,1)$ we get
\begin{equation}
{\cal N}_3=\rv{2\beta = |{\bf K}_- - {\bf K}_+|} \label{MKK0}
\end{equation}
(see { Appendix C}).
This gives the conventional expression for the Hall conductivity according to Eq. (\ref{HALL_4D}):
$$
\sigma_{xy} = \cor{+}{  \frac{{\cal N}_3}{4\pi^2}  }
            = \cor{-}\frac{|{\bf K}_+ - {\bf K}_-|}{4\pi^2}
$$
\rv{Notice, that we define here the conductivity $\sigma_{xy}$ through relation $j_x = - \sigma_{xy} E_y$. The sign minus originates from the two - dimensional notation relation between the conductivity tensor and resistivity $\rho_{xy}$:
$$
\left(\begin{array}{cc} 0 & -\sigma_{xy} \\ \sigma_{xy} & 0\end{array}\right) = \left(\begin{array}{cc} 0 & \rho_{xy} \\  - \rho_{xy} & 0\end{array}\right)^{-1}
$$
that gives $\sigma_{xy} = 1/\rho_{xy}$.}

If $m$
is larger than $1$, then ${\cal N}_3=0$. On the other hand, when $m$ approaches $-1$,
the two Fermi points tend to $\pm \pi$ (which represent actually the same point).
At $m=-1$ the value of ${\cal N}_3$ achieves its maximum equal to $2\pi$.
When $m$ is decreased further into the interval $(-3,-1)$, the new positions of the Fermi points
are
$${\bf K}^\prime_\pm =(0,\pi,\pm \beta^\prime,0);\,{\bf\tilde K}^{\prime}_\pm = (\pi,0,\pm\beta^\prime,0)$$
with $\beta'={\rm arccos}(m+2)$.
${\bf K}^\prime_+, {\bf \tilde K}_+^\prime$ are the right - handed Weyl points
while ${\bf K}^\prime_-,{\bf\tilde K}_-^\prime$ are the left - handed ones.
Using the machnery described in Appendix C, we come to the conclusion that
\begin{equation}
{\cal N}_3 = \rv{-4\beta^\prime + 2\pi}
\end{equation}
When $m$ crosses the value $-1$, the two Fermi points of Eq. (\ref{Fermi_point0}) disappear and the new two pairs of the Fermi points appear. However, the value of ${\cal N}_3$ is changed continuously. Nevertheless, the conventional expression for the Hall conductivity corresponding to Eq. (\ref{MKK0}) is broken.  Instead we may represent the value of ${\cal N}_3$ as
\begin{equation}
{\cal N}_3 =\rv{ \Big({K}^\prime_{-,3} - {K}^\prime_{+,3}\Big) + \Big({\tilde K}^\prime_{-,3} - {\tilde K}^\prime_{+,3}\Big) + 2\pi} \label{MKKP}
\end{equation}
and obtain the unconventional expression for the Hall conductivity
\begin{equation}\label{MKKPs}
\cor{-}{\sigma}_{xy} = \frac{\Big({K}^\prime_{+,3} - {K}^\prime_{-,3}\Big) + \Big({\tilde K}^\prime_{+,3} - {\tilde K}^\prime_{-,3}\Big)}{4\pi^2}
- \frac{1}{2\pi a}.
\end{equation}
Here we restore in our expressions the lattice spacing $a$ and
take the distance between the two Fermi points
$ \Big({\bf K}^\prime_{-} - {\bf K}^\prime_{+}\Big)_{3}
= \Big({\bf \tilde K}^\prime_{-} - {\bf \tilde K}^\prime_{+}\Big)_{3} = -  2\beta^\prime$
corresponding to the part of the straight line connecting them.
Eq. (\ref{MKKP}) may be interpreted as the sum of three contributions:
the first two contributions correspond to the two pairs of the Weyl points
while the last one corresponds to the contribution of the pair that has disappeared.
If there would be no two new pairs, the system would become the topological insulator
possessing the AQHE with the value of ${\cal N}_3$ equal to $2\pi$.

The further decrease of $m$ leads to the disappearance of the two pairs
${\bf K}^\prime_\pm$, ${\bf \tilde K}_\pm^\prime$.
The new pair appears for $m\in (-5,-3)$:
$${\bf K}^{\prime \prime}_\mp =(\pi,\pi,\pm \beta^{\prime \prime},0)$$
The value of ${\cal N}_3$ reads:
\begin{equation}
{\cal N}_3 = \rv{2\pi - 2\pi - 2\pi + \Big({\bf K}^{\prime \prime}_{-} - {\bf K}^{\prime \prime}_{+}\Big)_{3}
= 2\beta^{\prime \prime} - 2\pi} \label{MKKP2}
\end{equation}
and gives
\rv{\begin{equation}
\cor{-}{\sigma}_{xy} = \frac{\Big({K}^{\prime \prime}_{+,3} - {K}^{\prime \prime}_{-,3}\Big)}{4\pi^2}
+ \frac{1}{2\pi a} \label{MKKPs2}
\end{equation}}
At $m=-5$ the system undergoes transition to the insulator state, and for $m < -5$
$$
{\cal N}_3 = \rv{2\pi -2\pi -2\pi +2\pi = 0}
$$
that is $\sigma_{xy} = 0$.

We conclude, that the sufficiently large values of $\lambda$ lead to the essential change of the effective mass parameter and thus to the transition into the semimetal states with different configurations of the Weyl points as well as the different expressions for the Hall conductivity.

\subsection{$3+1$ D Weyl semimetals in the presence of Coulomb interaction}

Let us now discuss the three - dimensional model of Weyl semimetal with Coulomb interactions. In particular, the above considered particular model of Weyl semimetals corresponds to the action
\begin{eqnarray}\label{action_Coulomb_3+1D_a}
S &=& \int d\tau \sum_{{\bf x}}\Big[\bar{\psi}\Big(i(i \partial_{\tau} - A_4(i\tau,{\bf x})) - H(-i\partial_{\bf x} )\Big)\psi
\Big]\nonumber \\ && + \int d\tau \sum_{{\bf x,x'}}\Big[\frac{1}{8\pi \alpha}\phi_{\bf x'} {\cal U}_{\bf x',x}  \phi_{\bf x}  - \bar{\psi}(\tau,{\bf x})\psi(\tau, {\bf x})\phi(\tau, {\bf x})\Big]
\end{eqnarray}
where $A_4$ is the Euclidean scalar potential that corresponds to external electric field while
\begin{equation} \label{H(p)_3+1D}
H({\bf p}) = {\rm sin}\,p_1\, \sigma^2 - {\rm sin}\, p_2 \, \sigma^1 - (m - {\rm cos}\,p_3 + \sum_{i=1,2}(1-{\rm cos}\,p_i)) \, \sigma^3
\end{equation}
with $m \in (-1,1)$. With the interactions neglected this system contains the two Fermi points
\begin{equation}\label{Fermi_point}
{\bf K}_{\mp}=(0,0,\pm \beta,0),\quad \beta={\rm arccos} \, m.
\end{equation}
The interactions correspond to matrix ${\cal U}_{\bf x', x}$ that is inverse to the matrix of lattice Coulomb potential ${\cal D}_{\bf x', x}$:
\begin{eqnarray}\label{bosonic_lattice_diff_3d}
\sum_{\bf x}{\cal U}_{\bf z,x}{\cal D}_{\bf x y}=\delta_{\bf z, y}
\end{eqnarray}
 ${\cal D}_{\bf x, x'}$ is caused by the photon exchange in medium, which in the leading order is reduced to the Coulomb interaction because of the relative smallness of the Fermi velocity in medium. In the leading order it is given by the continuum Coulomb potential $1/|{\bf x} - {\bf x}^\prime|$, where ${|{\bf x} - {\bf x}^\prime|}$ is the real distance between the lattice points $\bf x', x$. There are also certain corrections to this potential that will not be discussed here but that may be relevant for certain physical phenomena.
 We will need only, that the Fourier transform $D(p)$ of ${\cal D}$ obeys $D(p) =  D(-p)$. Then repeating all the steps of Sect. \ref{SectScalarA} we come to the conclusion that the Coulomb interactions between electrons in Weyl semimetals do not affect the expression for the Hall conductivity, at least, in the first order in the effective fine structure constant in medium $\alpha$, that is the Hall conductivity is given by the expression of \cite{Z2016_1} expressed through the two point Green function of the interacting system:
 \begin{equation}\label{HALL_4D6}
{j}^k_{Hall} = \cor{-}\frac{1}{4\pi^2}\,{\cal N}_l\,\epsilon^{kjl} E_j ,
\end{equation}
where
\begin{eqnarray}\label{M_4D2}
{\cal N}_l &=& \rv{- \frac{1}{24 \pi^2}}\epsilon^{ijkl}
 \int_{} d^4 p \, {\rm Tr}\, {\cal G}_\alpha  \frac{\partial {\cal G}_\alpha^{-1}}{\partial p_i}
                        \frac{\partial {\cal G}_\alpha}{\partial p_j} \frac{\partial {\cal G}_\alpha^{-1}}{\partial p_k}.
\end{eqnarray}

Here ${\cal G}_\alpha$ is the two point Green function with the Coulomb interactions taken into account. It has the form
\begin{eqnarray}\label{Green_W_express_Coulomb4D}
{\cal G}_{\alpha}(p) =[i \omega - H({\bf p}) - \alpha \Sigma(p)]^{-1}.
\end{eqnarray}
The self-energy function is given by
{
\begin{eqnarray}\label{self-energy_}
\Sigma(p) = -\int {\cal G}_{\alpha = 0}(q) D(p-q) \frac{d^4 q}{(2\pi)^4}
\end{eqnarray}   }


\subsection{Transitions between the states of Weyl semimetals with various patterns of the Fermi points}

For the above considered particular model of Weyl semimetals,
{ the action in Eq.(\ref{action_Coulomb_3+1D_a}) can be written in the form
\begin{eqnarray}\label{action_Coulomb_3+1D_b}
S &=& \int d\tau \Big[\sum_{{\bf x}}\bar{\psi}\Big(i(i \partial_{\tau} - A_4(i\tau,{\bf x})) - H(-i\partial_{\bf x} )\Big)\psi  \nonumber \\
&& -\alpha \sum_{{\bf x,x'}}\bar{\psi}(\tau,{\bf x})\psi(\tau, {\bf x})V({\bf x-x'})\bar{\psi}(\tau,{\bf x'})\psi(\tau, {\bf x'})\Big],
\end{eqnarray}  }
where $A_4$ is the Euclidean scalar potential that corresponds to the external electric field,
and $H(p)$ is given by Eq.(\ref{H(p)_3+1D}) with $m \in (-1,1)$.
With the interactions neglected this system contains the two Fermi points
\begin{equation}\label{Fermi_point4}
{\bf K}_{\mp}=(0,0,\pm \beta,0),\quad \beta={\rm arccos} \, m.
\end{equation}
The electric current is given by Eq. (\ref{HALL_4D}) with ${\cal N}_3$ given by Eq. (\ref{M_4D2}).
The Green function entering this expression is
given by Eq.(\ref{Green_W_express_Coulomb4D}).
But the self-energy function is
\begin{eqnarray}\label{self-energy_4}
\Sigma(p) = -\int {\cal G}_{\alpha = 0}(q) \tilde{V}(p-q) {    \frac{d^4 q}{(2\pi)^4}   }
\end{eqnarray}
in which the Coulomb interaction in momentum space is
\begin {eqnarray}\label{bosonic_Green4}
\tilde{V}(p) =\sum_{\bf x} \frac{e^{i{\bf p\cdot x}}}{\sqrt{x_1^2+x_2^2+x_3^2}}.
\end{eqnarray}
The leading order of the self-energy function is given by
\begin{eqnarray}\label{self-energy4}
&& \Sigma^{(0)}(x,p)  = - \int \frac{1}{i q_4 - H({\bf q})} \tilde{V}(p-q) \frac{d^4 q}{(2\pi)^4} 
\end{eqnarray}
 $\Sigma^{(0)}(x,p)$ depends on $p_1$, $p_2$ and $p_3$.
The Coulomb interactions will change the value of Hall conductivity
through the self energy function. To see this more explicitly, let us denote
\begin{eqnarray}\label{Green_factorization}
G^{-1}=i\sigma^3(\sigma^k g_k(p)-ig_4(p))
\end{eqnarray}
and  represent the Hall conductivity as follows
(Eq.(55) in \cite{Z2016_1}):
\begin {eqnarray}\label{Hall_cond_4D}
\sigma_H =-\frac{1}{4\pi^2} {\cal N}_3
\end{eqnarray}
where
\begin{eqnarray}\label{N3_4D}
{\cal N}_3 &=&  \rv{-\frac{1}{2}}\sum_l \, \int_{y^{(l)}} {\rm sign}(g_4(y^{(l)})) \,{\rm Res}\,(y^{(l)}) dp_3
\end{eqnarray}
$y^{(l)}(s)$ is the line in 4D momentum space, where $g_1=g_2=g_3=0$, and
\begin{eqnarray}
{\rm Res}\,(y) &=&  \frac{1}{8 \pi} \epsilon^{ijk}\, \int_{\partial \Omega(y)} \, v_i  d v_j \wedge d v_k
\end{eqnarray}
(see Appendix C for details).

To the first order in $\alpha$, the Coulomb interactions do not change the positions of the lines of $y^{(l)}$, which are
given by
$$
y^{(1)}(p_3) = (p_1=0,p_2=0,p_3,p_4=0),$$$$ y^{(2)} (p_3) = (0,\pi, p_3,0),
$$$$ \quad y^{(3)}(p_3) = (\pi,0,p_3,0),\quad y^{(4)} (p_3) = (\pi,\pi,p_3,0),
$$
but it can change the position of the point when $g_4=0$ on the line, where  $g_4$ changes its sign. Therefore,
it changes the value of  ${\cal N}_3$ in Eq.(\ref{N3_4D}).

Let us take $y^{(1)}$ as an example. We represent
$\Sigma(p)=-i\sigma^3(\sigma^k f_k(p_1,p_2,p_3|m)-if_4(p_1,p_2,p_3|m))$.
{ Then $g_4$ in Eq.(\ref{Green_factorization}) is $g_4(y^{(1)})=m-{\rm cos} p_3 -\alpha \Sigma(y^{(1)})$,  }
$\Sigma(y^{(1)}) = - f_4(0,0,p_3|m)\sigma^3$ and
\begin{eqnarray}
&& f_4(0,0,p_3|m)=\nonumber\\
&& = \int \frac{m + 2-{\rm cos}\,q_1-{\rm cos}\,q_2-{\rm cos}\,q_3}
{\sqrt{{\rm sin}^2q_1 + {\rm sin}^2 q_2 + (m + 2-{\rm cos}q_1-{\rm cos}q_2-{\rm cos}q_3)^2}}\nonumber\\
&&
\tilde{V}(q_1,q_2,p_3-q_3) \frac{d^3 q}{2(2\pi)^3}.
\end{eqnarray}
The zero point of  { the function $g_4(y^{(1)})=g_4(0,0,p_3)$   }
is the solution of equation  $$g_4(0,0,p_3)=m-{\rm cos}\,p_3+\alpha  f_4(0,0,p_3|m)=0.$$
Its solution is  $p_3={\rm arccos}\,(m+\alpha f_4(0,0,\beta|m))+O(\alpha^2)$.
The value of ${\cal N}_3 $ is modified (it is equal to \rv{$2{\rm arccos}\,(m)$} in the absence of Coulomb interactions):
${\cal N}_3 =\rv{2{\rm arccos}\,(m+\alpha f_4(0,0,\beta|m))}$, which is still equal to the difference between the positions of the Weyl points. Their positions are modified due to the interactions.

Besides, the pattern of the AQHE may be modified in a more complicated way.
Namely, without interactions for $-1<m<1$,  the signs of $g_4$ along
the lines $y^{(2)}$, $y^{(3)}$ and $y^{(4)}$ remain positive.
However, say, if $m$ remains slightly larger than $-1$,
the Coulomb interactions may lead to the appearance of the zero of
{ $g_4(y^{(2)})=g_4(0,\pi,p_3)$ } that is the solution of equation:
$$g_4 (0,\pi,p_3)=2+m-{\rm cos}\,p_3+\alpha  f_4(0,\pi,p_3|m)=0$$
This equation may have a solution $p_3=\beta^{\prime}$ if $f_4(0,\pi,p_3|m)$ is able to become negative at small $p_3$.
Our numerical calculations on a $40\times 40\times 40$ lattice show that
{ $f_4(0,\pi,0|m=-1)\approx -0.15$}. Therefore, the zero of $g_4$ really may appear
along the line $y^{(2)}$ (near $p_3=0$), when parameter $m$ is close to $-1$.
The new Fermi points will be located at
$${\bf K}_\pm =(0,\pi,\pm \beta^{\prime},0);\,{\bf K}^\prime_\pm = (\pi,0,\pm\beta^{\prime},0)$$
On the other hand, if in addition $f_4(0,0,p_3|m)$ is able to become negative
at $p_3$ close to $\pi$, then the zero of $g_4(0,0,p_3)$ may disappear together with the conventional pair of Weyl points of Eq. (\ref{Fermi_point4}).
Our numerical results show that
{$f_4(0,0,\pi|m=-1)\approx -1.4$}. Therefore, this may really occur.

Since $f_4(0,0,\pi|m=-1)<f_4(0,\pi,0|m=-1)<0$, the interval $m\in (-1,1)$ may be
divided into 3 regions (when $\alpha <<1$):

(I) $-1-\alpha f_4(0,0,\pi|m=-1)<m<1 $: $g_4$ has one zero along the line $y^{(1)}$,
but there is no zero along $y^{(2)}$, and $y^{(3)}$. This case is similar to the original
one without interaction.

(II) $-1-\alpha f_4(0,\pi,0|m=-1)< m < -1-\alpha f_4(0,0,\pi|m=-1)$:
$g_4$ does not have zero along $y^{(1)}$, $y^{(2)}$, and $y^{(3)}$.
Therefore,
\rv{\begin{equation}
{\sigma}_{xy} =  \cor{+}\frac{1}{2\pi a}.
\end{equation}}
This is the case, when interactions bring the system to the insulator phase. The insulator has a nontrivial topology that causes the AQHE according to the mechanism discussed in Sect. 9 of \cite{Z2016_1}.

(III) $-1<m<-1-\alpha f_4(0,\pi,0|m=-1)$:
$g_4$ does not have zero along $y^{(1)}$, but has one zero
along each of the lines $y^{(2)}$ and $y^{(3)}$.

The Hall conductivity is then given by
\begin{equation}
\cor{-}{\sigma}_{xy} = \frac{\Big({K}_{+,3} - {K}_{-,3}\Big) + \Big({K}^\prime_{+,3} - { K}^\prime_{-,3}\Big)}{4\pi^2} - \frac{1}{2\pi a} \label{MKKPs4}
\end{equation}
as in the case of the four - fermion interactions discussed above (see Eq. (\ref{MKKPs}).

\section{Conclusions and discussion}

In the present paper we investigated the influence of interactions on the AQHE. \revv{We consider the particular tight - binding models of the $2+1$ D topological insulator discussed in \cite{tb2d,tb2d2,tb2d3,Z2016_1} and of the $3+1$ D Weyl semimetals (discussed, for example, in \cite{tb1,tb2,tb3,tb4,tb5}). } We consider the effect of the four - fermion interactions, Yukawa and Coulomb interactions.
As expected, we obtain, that in all considered cases the AQHE conductivities are given by expressions discussed in \cite{Z2016_1}. Those expressions are composed of the two point Green functions (defined in momentum space) of the interacting systems. For the case of the insulators the given expressions are the topological invariants and are not changed when the system is modified smoothly. For the case of Weyl semimetals strictly speaking the obtained expressions are not the topological invariants. The smooth modification of the system may lead to the smooth change of their values. However, certain features are inherited by these expressions from the case of the topological insulators.  At least, up to the one - loop order the corrections to the conductivities due to interactions may be taken into account only through the modification of the mentioned one - particle Green functions. The other corrections do not appear. We expect that this result remains valid to all orders of the perturbation expansion (at least, for the sufficiently small values of coupling constants).
We summarize our main results as follows.

In the case of the 2+1 D topological insulators the four - fermion interactions at the one-loop level
renormalize the mass  parameter of the considered tight - binding models.
If the strength of the interaction is large enough,
the system can be driven to the phase with the value of the Hall conductivity
different from that of the non - interacting model.
Otherwise, if  the strength is lower than a certain threshold, the
Hall conductivity remains the same.
The effects of Yukawa and Coulomb interactions are similar.
To the best of our knowledge for the considered models the perturbative corrections to Hall conductivity are considered for the first time. Moreover, corrections due to Yukawa interactions are considered to all orders in perturbation theory. (This consideration may be extended easily to exchange by any bosonic excitations.) In the latter consideration
we used an original method to prove diagrammatically that in the presence of interactions the Hall conductivity is given by the topological quantity expressed through the full fermion propagator.
 The essence of our proof is to construct the bubble-like
Feynman diagram (progenitor) for a group of diagrams, which contribute to the
Hall current. This  method is somehow similar to the progenitor approach
used by Coleman and Hill in QED3 \cite{parity_anomaly}. However, unlike \cite{parity_anomaly} we consider the more complicated model without relativistic invariance. It is worth mentioning that the different diagrammatic method of \cite{Matsuyama:1986us,Imai:1990zz} (applied originally to the QHE in the presence of magnetic field) may, in principle, be extended to the intrinsic AQHE. However, this remains out of the scope of the present paper.

The situation for the $3+1$D Weyl semimetal is different.
Due to the four - fermion interactions, the   $3+1$ D Weyl semimetal may drop to the different state of the semimetal with the unconventional expression for the AQHE conductivity (the conventional AQHE conductivity in the Weyl semimetal is proportional to the distance between the Weyl points of opposite chiralities). This is not a topological phase transition. However, the Weyl semimetal may also undergo the true topological phase transition to the insulator phase.
Coulomb interactions  affect the systems more strongly. Not only the parameters are renormalized, the very dependence of the propagators on momenta becomes different. the Coulomb interactions also may bring the $3+1$  D Weyl semimetals to the states with the unconventional expression for the Hall conductivity. In the latter case we may deal both with the true phase transition to the topological insulator exhibiting the AQHE and with the smooth modification of the Weyl point pattern that gives rise to the unconventional AQHE conductivity.
To the best of our knowledge, we considered {Coulomb interaction} corrections to the Hall conductivity of Weyl semimetals for the first time.

The authors kindly acknowledge useful discussions with I.Fialkovsky, Xi Wu, and M.Suleymanov.

\label{SectConcl}

\section*{Appendix A. AQHE conductivity in the noninteracting lattice model}

\label{Sect2}

Here we repeat for the completeness some of the results reported earlier in \cite{Z2016_1,SZ2018,ZW2019} that are used in the main text of the present paper. We  consider the lattice $2+1$ D tight - binding fermionic model with the following partition function
\rv{\begin{eqnarray}
Z&=&\int {D\bar{\Psi} D\Psi } exp \Bigl(\int d\tau d\tau^\prime\sum_{{\bf r}_n,{\bf r}_m}\bar\Psi^T({\bf r}_m,\tau)\nonumber\\&&\left(-i\mathcal{D}(\tau,\tau^\prime)_{{\bf r}_n,{\bf r}_m}\right)\Psi({\bf r}_n,\tau^\prime)\Bigr)
\label{Z00} \end{eqnarray}}
$\mathcal{D}_{\bf x,y}(\tau,\tau^\prime)$ depends on $2D$ lattice sites $\bf x$, $\bf y$ and on the points along the axis of imaginary time $\tau,\tau^\prime$. The lattice is assumed to be rectangular for simplicity. Variables $\Psi,\bar{\Psi}$ are the anti - commuting Grassmann - valued fields. We may rewrite this partition function in momentum space:
\begin{eqnarray}\label{Z01}
Z = \int {D\bar{\psi} D\psi } \, {\rm exp}\Big(  \int_{\cal M} \frac{d^3 {p}}{(2\pi)^3}\bar{\psi}^T({p}){Q}(p)\psi({p}) \Big),
\end{eqnarray}
 Here integration is over the fields $\bar{\psi}$ and $\psi$ that are functions of momenta. The partition function of Eq. (\ref{Z01}) allows to describe the non - interacting fermions.  $Q(p)$ is the Fourier transform of $D_{\bf x y}(\tau,\tau^\prime)$ that has the meaning of the inverse fermion propagator.

External gauge field $A(x)$ may be introduced to the model through the Peierls substitution  \cite{Z2016_1,SZ2018}:
\begin{eqnarray}\label{Z01_a}
Z &=& \int { D\bar{\psi}D\psi} \, {\rm exp}\Big(  \int \frac{d^D {p}}{(2\pi)^3}\bar{\psi}^T({p}){Q}(p - A(i\partial_p))\psi({p}) \Big).
\end{eqnarray}
We assume in this expression the symmetrization of the products of operators within ${\cal Q}(p - A(i\partial_p))$.

The matrix elements of $\hat{Q} = Q(p-A(i\partial_p))$ and its inverse $\hat{G} = \hat{Q}^{-1}$ are denoted here by ${\cal Q}(p,q)$ and ${\cal G}(p,q)$ correspondingly:
$$
{\cal Q}(p,q) = \langle p|\hat{Q}| q\rangle, \quad {\cal G}(p,q) = \langle p|\hat{Q}| q\rangle
$$
It is assumed that the basis elements $|q\rangle$ of the space of functions are normalized as $\langle p| q\rangle = \delta^{(3)}(p-q)$. We also have
$$
\langle p|\hat{Q}\hat{G}|q\rangle = \delta({ p} - {q}).
$$
Eq. (\ref{Z01_a}) may be rewritten as follows
\rv{\begin{eqnarray}
Z = \int {D\bar{\psi}D\psi }\, {\rm exp}\Big(  \int \frac{d^D {p}_1}{\sqrt{(2\pi)^3}} \int \frac{d^3 {p}_2}{\sqrt{(2\pi)^3}} \bar{\psi}^T({p}_1){\cal Q}(p_1,p_2)\psi({p}_2) \Big),\label{Z1}
\end{eqnarray}}
while the propagator of fermion is given by
\rv{\begin{eqnarray}
{\cal G}_{ab}(k_1,k_2)&=& \frac{1}{Z}\int D\bar{\psi}D\psi \, {\rm exp}\Big(  \int \frac{d^3 {p}_1}{\sqrt{(2\pi)^3}} \int \frac{d^3 {p}_2}{\sqrt{(2\pi)^3}}\nonumber\\&&\bar{\psi}^T({p}_1){\cal Q}(p_1,p_2)\psi({p}_2) \Big) \frac{\bar{\psi}_b(k_1) }{\sqrt{(2\pi)^3}} \frac{{\psi}_a(k_2)}{\sqrt{(2\pi)^3}}\label{G1}
\end{eqnarray}}
Components of $\psi$ are enumerated by $a,b$. Below we omit those indices.

Wigner transformation of $\cal G$ is equal to the Weyl symbol of $\hat{G}$:
\begin{eqnarray}
{G}_W(x,p) \equiv \int dq e^{ix q} {\cal G}({p+q/2}, {p-q/2})\label{GWx}
\end{eqnarray}
The Groenewold equation \cite{Z2016_1} relates them as follows
\begin{eqnarray}
{G}_W(x,p) \star Q_W(x,p) = 1 \label{Geq}
\end{eqnarray}
that is
\begin{eqnarray}
&1 =
{G}_W(x,p)
e^{\frac{i}{2} \left( \overleftarrow{\partial}_{x}\overrightarrow{\partial_p}-\overleftarrow{\partial_p}\overrightarrow{\partial}_{x}\right )}
Q_W(x,p)
\label{GQW}
\end{eqnarray}
By $x$ we denote the coordinates that are composed of the discrete lattice coordinates and the continuous imaginary time. The differentiation over $x$ may be defined if the definition of the functions of coordinates is extended to continuous values. Eq. (\ref{GQW}) is valid as long as the variation  of the field $A(x)$ on the distance of the order of the lattice spacing may be neglected.

For an arbitrary lattice model the calculation of Weyl symbol $Q_W$ of  $\hat{Q}=Q(p-A(i\partial_p))$ is technically complicated. This problem has been solved in \cite{SZ2018} for the particular case of lattice Wilson fermions. In general case if the field $A$ is slowly varying, then up to the terms linear in the field strength ${Q}_W(p,x)= Q_W(p-A(x)) \equiv Q(p-A(x))$ (see \cite{Z2016_1}).

Electric current may be expressed as follows
\rv{\begin{eqnarray}
J^k(x) &=& -\int d^3 p\, {\rm Tr} \, G_W(p,x)  \partial_{p_k} Q_W(p - { A}(x))  \label{J3}
\end{eqnarray}}

The solution of Groenewold equation may be obtained via the expansion in powers of derivatives. The first two terms in this expansion are given by \cite{SZ2018}:
\begin{eqnarray}
G_W(x,p)  &= &G^{(0)}_W(x,p) +  G^{(1)}_W(x,p) + ...  \label{Gexp}\\
G^{(1)}_W  &= &\cor{+}\frac{i}{2}  G^{(0)}_W \frac{\partial \Big[ G^{(0)}_W\Big]^{-1}}{\partial p_i}  G^{(0)}_W  \frac{\partial  \Big[ G^{(0)}_W\Big]^{-1}}{\partial p_j}  G^{(0)}_W
A_{ij} (x)\nonumber
\end{eqnarray}
Here $G^{(0)}_W(x,p)$ is defined as the Green function with the field strength $A_{ij} = \partial_i A_j - \partial_j A_i$ neglected. It is given by
\begin{eqnarray}
&& G^{(0)}_W(x,p)  = Q^{-1}(p-A(x))\label{Q0}
\end{eqnarray}
Substituting Eq. (\ref{Gexp}) to Eq. (\ref{J3}) gives
the following result for the part of the current proportional to $A_{ij}$:
\begin{equation}
\langle J^{k} \rangle \approx \cor{-} \frac{1}{4\pi}{\cal M}^{ijk} A_{ij}
\end{equation}
Tensor ${\cal M}^{ijk}$ is the topological invariant in momentum space given by
\begin{eqnarray}
{\cal M}^{ijk} &= & \epsilon^{ijk} {\cal M},\quad  {\cal M} = \int \,{\rm Tr}\, \nu_{} \,d^3p \label{j2d00}\\ \nu_{} & = & \rv{ \frac{i}{3!\,4\pi^2}\,\epsilon_{ijk}\, \Big[  {Q}(p) \frac{\partial {Q}^{-1}(p)}{\partial p_i} \frac{\partial  {Q}( p)}{\partial p_j} \frac{\partial  {Q}^{-1}(p)}{\partial p_k} \Big] }\nonumber
\end{eqnarray}

\section*{Appendix B. Calculation of ${\cal N}$ for the $2+1$ D systems}

Here we repeat the calculation presented in Appendix C of \cite{Z2016_1}. We calculate  the topological invariant ${\cal N}$  in the case, when the Green function has the form
 \begin{equation}
 {\cal G}^{-1}(p) = i\sigma^3\Big(\sum_{k}\sigma^{k} g_{k}(p) - i g_4(p)\Big)\label{G12d}
 \end{equation}
 where $\sigma^k$ are Pauli matrices while $g_k(p)$ and $g_4(p)$ are the real - valued functions, $k = 1,2,3$. Let us define
 \begin{equation}
 {\cal H}(p) = \Big(\sum_{k}\sigma^{k} \hat{g}_{k}(p) - i \hat{g}_4(p)\Big)
 \end{equation}
where $\hat{g}_k = \frac{g_k}{g}$, and $g = \sqrt{\sum_{k=1,2,3,4}g_k^2}$. Then
\rv{\begin{eqnarray}
{\cal N} &=&  -\frac{1}{24 \pi^2} {\rm Tr}\, \int_{} \, {\cal G}^{-1} d {\cal G} \wedge d {\cal G}^{-1} \wedge d {\cal G}   \nonumber\\
&=& -\frac{1}{24 \pi^2} {\rm Tr}\, \int_{} \, {\cal H} d {\cal H}^+ \wedge d {\cal H} \wedge d {\cal H}^+\nonumber\\
&=&  -\frac{1}{48 \pi^2} {\rm Tr}\,\gamma^5 \int_{} \, \tilde{\cal H} d \tilde{\cal H} \wedge d \tilde{\cal H} \wedge d \tilde{\cal H}\label{N3AH}
\end{eqnarray}}
where
\begin{equation}
\tilde{\cal H}(p) = \sum_{k=1,2,3,4}\gamma^{k} \hat{g}_{k}(p) = i{\rm diag}\,({\cal H},-{\cal H}^+)\gamma^4
\end{equation}
and $\gamma^k$ are Euclidean Dirac matrices in chiral representation, that is
{
\begin{eqnarray}
\gamma^i=\left(\begin{array}{cc} 0 & i\sigma^i \\ -i\sigma^i & 0 \end{array}\right),
\quad
\gamma^4=\left(\begin{array}{cc} 0 & I_2 \\ I_2 & 0 \end{array}\right)
\end{eqnarray}
with $i=1,2,3$.}
$\gamma^5$  in chiral representation is given by ${\rm diag}(1,1,-1,-1)$. This gives
\begin{eqnarray}
{\cal N} &=& \rv{ \frac{1}{12 \pi^2}} \epsilon^{ijkl}\, \int_{} \, \hat{g}_i d \hat{g}_j \wedge d \hat{g}_k \wedge d \hat{g}_l
\end{eqnarray}
Let us introduce the parametrization
\begin{equation}
\hat{g}_4 = {\rm sin}\,\alpha, \quad \hat{g}_i = k_i\,{\rm cos}\,\alpha
\end{equation}
where $i=1,2,3$ while $\sum_{i}k^2=1$, and $\alpha \in [-\pi/2,\pi/2]$. Let us suppose, that $\hat{g}_4(p)=0$ on the boundary of momentum space $p\in \partial {\cal M}$. This gives
 \rv{ \begin{eqnarray}
{\cal N} &=&  \frac{1}{4 \pi^2} \epsilon^{ijk}\, \int_{\cal M} \, {\rm cos}^2 \alpha\, k_i\,  d\,\alpha  \wedge d k_j \wedge d k_k\\
&=&  \frac{1}{4 \pi^2} \epsilon^{ijk}\, \int_{\cal M} \, k_i\,  d(\alpha/2+\frac{1}{4}{\rm sin}\,2\alpha)  \wedge d k_j \wedge d k_k \nonumber\\
&=& -\sum_l \frac{1}{4 \pi^2} \epsilon^{ijk}\, \int_{\partial{\Omega(y_l)}} \, k_i\,  (\alpha/2+\frac{1}{4}{\rm sin}\,2\alpha)   d k_j \wedge d k_k\nonumber
\end{eqnarray}}
In the last row $\Omega(y_l)$ is the small vicinity of point $y_l$ of momentum space, where vector $k_i$ is undefined. The absence of the singularities of $\hat{g}_k$ implies, that $\alpha \rightarrow \pm\pi/2$ at such points.

This gives
\begin{eqnarray}
{\cal N} &=& \rv{-  \frac{1}{2}}\sum_l \, {\rm sign}(g_4(y_l)) \,{\rm Res}\,(y_l)
\end{eqnarray}
We use the notation:
\begin{eqnarray}
{\rm Res}\,(y) &=&  \frac{1}{8 \pi} \epsilon^{ijk}\, \int_{\partial \Omega(y)} \, k_i  d k_j \wedge d k_k
\end{eqnarray}
It is worth mentioning, that this symbol obeys { $\sum_l {\rm Res}\,(y_l)=0$}.

Let us illustrate the above calculation by the consideration of the  particular example of the system with the Green function ${\cal G}^{-1} = i \omega - H(p)$, where the Hamiltonian has the form
\begin{equation}
H = {\rm sin}\,p_1\, \sigma^2 - {\rm sin}\, p_2 \, \sigma^1 - (m + \sum_{i=1,2}(1-{\rm cos}\,p_i)) \, \sigma^3
\end{equation}
This gives
\begin{equation}
-i\sigma^3{\cal G}^{-1} = {\rm sin}\,p_1\, \sigma^1 + {\rm sin}\, p_2 \, \sigma^2 +  \omega \, \sigma^3 - i (m + \sum_{i=1,2}(1-{\rm cos}\,p_i))
\end{equation}
The boundary of momentum space corresponds to $\omega = \pm \infty$. We have
$$\hat{g}_4(p) = \frac{m  + \sum_{i=1,2}(1-{\rm cos}\,p_i)}{\sqrt{(m + \sum_{i=1,2}(1-{\rm cos}\,p_i))^2+ {\rm sin}^2\,p_1+ {\rm sin}^2\,p_2 + \omega^2}} $$
For example, for $m \in (-2,0)$ we have
\begin{eqnarray}
\hat{g}_4(p) & = & 0, \quad p\in \partial{\cal M}\nonumber\\
\hat{g}_4(p) & = & -1, \quad \hat{g}_i(p)  = 0\quad ({i} = 1,2,3),\quad  p = (0,0,0)\nonumber\\
\hat{g}_4(p) & = & 1,\quad  \hat{g}_i(p)  = 0\quad (i = 1,2,3),\quad  p = (0,\pi,0) \nonumber\\
\hat{g}_4(p) & = & 1,\quad  \hat{g}_i(p)  = 0\quad (i = 1,2,3),\quad  p = (\pi,0,0)\nonumber\\
\hat{g}_4(p) & = & 1,\quad  \hat{g}_i(p)  = 0\quad (i = 1,2,3),\quad  p = (\pi,\pi,0)
\end{eqnarray}
Therefore, we get immediately
\rv{\begin{eqnarray}
{\cal N} &=&  \frac{1}{2} - \frac{1}{2} (-1)-\frac{1}{2}(-1) - \frac{1}{2} =  1
\end{eqnarray}}
In the similar way \rv{${\cal N}  = -1 $} for $m \in (-4,-2)$ and ${\cal N}  = 0 $ for $m \in (-\infty,-4)\cup (0,\infty)$.

\section*{Appendix C. AQHE in the considered toy model of $3+1$ D  Weyl semimetal}
\label{SectWeylSem}

Here we repeat the calculation of the AQHE conductivity given in Sect. 7.1 and Sect. 10.1 of \cite{Z2016_1}. The considered toy model of Weyl semimetal corresponds to the Green function ${\cal G}^{-1} = i \omega - H(p)$ and the Hamiltonian of the form
\begin{equation}
H = {\rm sin}\,p_1\, \sigma^2 - {\rm sin}\, p_2 \, \sigma^1 - (m^{}-{\rm cos}\,p_3 + \sum_{i=1,2}(1-{\rm cos}\,p_i)) \, \sigma^3
\end{equation}
For $m\in (0,1)$ the system contains the two Fermi points
\begin{equation}
{\bf K}_\pm = (0,0,\pm \beta,0), \quad \beta = {\rm arccos}\,m
\end{equation}
Although the Green function contains singularities, the integral in the expression for ${\cal N}_3$ is convergent. We first integrate over momentum space with the small vicinities of the poles subtracted, and then consider the limit, when those vicinities are infinitely small.
We represent the Green function as follows
\begin{eqnarray}
-i\sigma^3{\cal G}^{-1} &=& {\rm sin}\,p_1\, \sigma^1 + {\rm sin}\, p_2 \, \sigma^2 +  \omega \, \sigma^3 \nonumber\\&&- i (m - {\rm cos}\,p_3+ \sum_{i=1,2}(1-{\rm cos}\,p_i))
\end{eqnarray}
The Hall current is given by
\begin{equation}
{j}^k_{Hall} = \cor{-}\frac{1}{4\pi^2}\,{\cal N}_l\,\epsilon^{kjl}E_j,\label{HALLj3d}
\end{equation}
where ${\cal N}_l$ is given by
\begin{eqnarray}
{\cal N}_l &=&  \rv{-\frac{1}{3!\,4\pi^2}}\,\epsilon_{ijkl}\,\int_{} \,\,d^4p\,{\rm Tr} \Big[  {\cal G} \frac{\partial {\cal G}^{-1}}{\partial p_i} \frac{\partial  {\cal G}}{\partial p_j} \frac{\partial  {\cal G}^{-1}}{\partial p_k} \Big]  \label{nuGHall}
\end{eqnarray}

Let us denote
 \begin{equation}
 {\cal G}^{-1}({ p}) = i\sigma^3\Big(\sum_{k}\sigma^{k} g_{k}({ p}) - i g_4({ p})\Big)\label{G12d2}
 \end{equation}
and $\hat{g}_k = \frac{g_k}{g}$, $g = \sqrt{\sum_{k=1,2,3,4}g_k^2}$. Next, we  introduce the parametrization
\begin{equation}
\hat{g}_4 = {\rm sin}\,\alpha, \quad \hat{g}_a = k_a\,{\rm cos}\,\alpha
\end{equation}
where $a=1,2,3$ while $\sum_{a=1,2,3}k^2_a=1$, and $\alpha \in [-\pi/2,\pi/2]$. One can check that $\hat{g}_4({ p})=0$ on the boundary of momentum space. This gives
 \rv{ \begin{eqnarray}
&&{\cal N}_n =  -\frac{1}{4 \pi^2} \epsilon^{abc}\,\epsilon^{ijkn} \int_{\cal M} \, {\rm cos}^2 \alpha\, k_a\,  \partial_i\alpha \, \partial_j k_b\, \partial_k k_c\, d^4p \nonumber\\ &&= - \frac{1}{4 \pi^2} \epsilon^{abc}\, \int_{\cal M} \, k_a\,  d(\alpha/2+\frac{1}{4}{\rm sin}\,2\alpha)  \wedge d k_b \wedge d k_c\wedge dp_n \nonumber\\ &&= \sum_l \frac{1}{4 \pi^2} \epsilon^{abc}\, \int_{\partial{\Omega(y_l)}} \, k_a\,  (\alpha/2+\frac{1}{4}{\rm sin}\,2\alpha)   d k_b \wedge d k_c \wedge dp_n\nonumber
\end{eqnarray}}
Now $\Omega(y^{(l)})$ is the small vicinity of line $y^{(l)}(s)$ in momentum space, where vector $k_i$ is undefined. Along these lines $\alpha \rightarrow \pm\pi/2$.
We have
 \rv{ \begin{eqnarray}
{\cal N}_j &=&   \frac{1}{2}\sum_l \, \int_{y^{(l)}(s)} \, {\rm sign}(g_4(y^{(l)})) \,{\rm Res}\,(y^{(l)})\wedge dp_j\label{calMproj}
\end{eqnarray}}
Here we denote
\begin{eqnarray}
{\rm Res}\,(y^{(l)}) &=&  \frac{1}{8 \pi} \epsilon^{ijk}\, \int_{\Sigma(y^{(l)})} \, k_i  d k_j \wedge d k_k
\end{eqnarray}
 the corresponding integral is  along the  infinitely small surface $\Sigma$, which is wrapped around the line $y^{(j)}(s)$ near to the given point of this line.
\rv{Notice, that in general case ${\rm Res}\,(y^{(l)})$ enters the expression for the differential form. However, if we chose $\Sigma$ that belongs to three - dimensional hypersurface orthogonal to the curve $y^{(j)}(s)$, then ${\rm Res}\,(y^{(l)})$ and $dp_j$ are factorized. ${\rm Res}\,(y^{(l)})$ becomes the integer number, and we get
\begin{eqnarray}
{\cal N}_j &=&   \frac{1}{2}\sum_l \, \int_{y^{(l)}(s)} \, {\rm sign}(g_4(y^{(l)})) \,{\rm Res}\,(y^{(l)})\times dp_j \,(-1)^{N[\Sigma;y^{(j)}(s)]}\label{calMproj2}
\end{eqnarray}
Here $(-1)^{N[\Sigma;y^{(j)}(s)]}  = \pm 1$ depending on the mutual orientation of the surface $\Sigma$ and the curve $y^{(j)}(s)$. For example, if $\Sigma$ belongs to the hyperplane $(x_1,x_2,x_4)$ while the line $y^{(j)}(s)$ coincides with the third axis, then $(-1)^{N[\Sigma;y^{(j)}(s)]}=-1$.}

For the given particular Hamiltonian
$$\hat{g}_4(p) = $$ $$= \frac{m - {\rm cos}\,p_3 + \sum_{i=1,2}(1-{\rm cos}\,p_i)}{\sqrt{(m- {\rm cos}p_3 + \sum_{i=1,2}(1-{\rm cos}p_i))^2+ \sum_{i=1,2}{\rm sin}^2p_i + \omega^2}} $$
and
\begin{eqnarray}
\hat{g}_4(p) & = & 0, \quad p\in \partial{\cal M}\nonumber\\
\hat{g}_4(p) & = & -1, \quad \hat{g}_i(p)  = 0\quad ({i} = 1,2,3),\quad  p = (0,0,p_3,0),\nonumber\\ && {\rm for}\quad p_3\in (-\beta,\beta)
\nonumber\\
\hat{g}_4(p) & = & 1, \quad \hat{g}_i(p)  = 0\quad (i = 1,2,3),\quad  p = (0,0,p_3,0),\nonumber\\ && {\rm for}\quad p_3\in (-\pi,-\beta)\cup (\beta,\pi)\nonumber\\
\hat{g}_4(p) & = & 1,\quad  \hat{g}_i(p)  = 0\quad (i = 1,2,3),\quad  p = (0,\pi,p_3,0),\nonumber\\ && {\rm for}\quad p_3\in (-\pi,-\pi) \nonumber\\
\hat{g}_4(p) & = & 1,\quad  \hat{g}_i(p)  = 0\quad (i = 1,2,3),\quad  p = (\pi,0,p_3,0),\nonumber\\ && {\rm for}\quad p_3\in (-\pi,-\pi)\nonumber\\
\hat{g}_4(p) & = & 1,\quad  \hat{g}_i(p)  = 0\quad (i = 1,2,3),\quad  p = (\pi,\pi,p_3,0),\nonumber\\ && {\rm for}\quad p_3\in (-\pi,-\pi)\label{listzerosW1}
\end{eqnarray}
Therefore ${\cal N}_1 = {\cal N}_2 ={\cal N}_4 = 0$ while
\rv{\begin{eqnarray}
{\cal N}_3 &=&  -\frac{2\pi-2\beta}{2} + \frac{2\beta}{2} - \frac{2\pi}{2} (-1)-\frac{2\pi}{2}(-1) - \frac{2\pi}{2} \nonumber\\&=& 2\beta
\end{eqnarray}}
Thus we come to the expression for the AQHE current
\rv{\begin{equation}
{j}^k_{Hall} = \cor{-}\frac{\beta}{2\pi^2}\,\epsilon^{kj3}E_j,\label{HALLj3dp}
\end{equation}}
This result coincides with the one of the naive low energy effective field theory \cite{Zyuzin:2012tv}.

\end{document}